\documentclass[11pt]{article}
\pdfoutput = 1

\usepackage[utf8]{inputenc}
\usepackage{color,graphicx}
\usepackage{epsfig}
\usepackage{amsmath}
\usepackage{verbatim}
\usepackage{amssymb}
\usepackage{mathabx}
\usepackage[mathcal]{eucal}
\usepackage{stmaryrd}
\usepackage{esint}
\usepackage{braket}
\usepackage{physics}
\usepackage{amsfonts}
\usepackage{cite}
\usepackage{array}
\usepackage{setspace}
\usepackage{braket}
\usepackage{float}
\usepackage{url}
\usepackage{mathtools}
\usepackage{bbold}
\usepackage{slashed}
\usepackage{tikz}
\usepackage{graphicx}
\usepackage{epstopdf}
\usepackage{subfigure}
\usepackage[labelfont=bf,font={sf}]{caption}
\usepackage{pgfplots}
\usepackage{mathrsfs}
\usepackage{fancybox}
\usepackage{eurosym}
\usepackage{tcolorbox}
\usepackage{tensor}
\allowdisplaybreaks

\usepackage[margin = 2.2cm]{geometry}
\usepackage[ragged]{footmisc}
\usepackage[bookmarks=true,bookmarksnumbered=false,
hyperindex=true,bookmarksopen=true,hyperfigures=true,
colorlinks=true,linkcolor=webblue,citecolor=webgreen,urlcolor=webblue,breaklinks]{hyperref}
\definecolor{webgreen}{rgb}{0, 0.5, 0}
\definecolor{webblue}{rgb}{0, 0, 0.5}
\definecolor{webred}{rgb}{0.5, 0, 0}
\definecolor{darkgreen}{rgb}{0,0.5,0}

\def\bra#1{\langle #1 |}
\def\ket#1{| #1 \rangle}

\renewcommand{\d}{\mathrm{d}}
\renewcommand{\i}{\mathrm{i}}

\def\ben{\begin{equation}}
\def\een{\end{equation}}

\let\a=\alpha \let\b=\beta  \let\d=\delta 
   
     \let\r=v
\let\s=\sigma

\def\be{\begin{equation}}
\def\ee{\end{equation}}
\def\ba{\begin{array}}
\def\ea{\end{array}}

\def\dalemb#1#2{{\vbox{\hrule height .#2pt
       \hbox{\vrule width.#2pt height#1pt \kern#1pt
               \vrule width.#2pt}
       \hrule height.#2pt}}}

\newcommand{\bea}{\begin{eqnarray}}
\newcommand{\eea}{\end{eqnarray}}

\let\tilde=\widetilde

\renewcommand{\d}{\mathrm{d}}
\renewcommand{\i}{\mathrm{i}}
\renewcommand{\S}{\textsf{S}_0}

\numberwithin{equation}{section}

\title{}

\pgfplotsset{compat=1.18} 
\begin{document}

\thispagestyle{empty}

\begin{center}
~\vspace{1cm}

{\LARGE \bf 
  Firewalls at exponentially late times
}
    
\vspace{0.4in}
    
{\bf Andreas Blommaert$^1$, Chang-Han Chen$^{2,3}$ and Yasunori Nomura$^{2,3,4,5}$}

\vspace{0.25in}

{
$^1$SISSA and INFN, Via Bonomea 265, 34127 Trieste, Italy\\
$^2$Berkeley Center for Theoretical Physics, Department of Physics, \\
University of California, Berkeley, CA 94720, USA\\
$^3$ Theoretical Physics Group, Lawrence Berkeley National Laboratory, Berkeley, CA 94720, USA\\
$^4$ RIKEN iTHEMS, Wako, Saitama 351-0198, Japan\\
$^5$ Kavli IPMU (WPI), UTIAS, The University of Tokyo, Kashiwa, Chiba 277-8583, Japan
}

\vspace{0.1in}
    
{\tt ablommae@sissa.it, changhanc@berkeley.edu, ynomura@berkeley.edu}
\end{center}

\vspace{0.3in}

\begin{abstract}
\noindent 
We consider a version of the typical state firewall setup recently reintroduced by Stanford and Yang, who found that wormholes may create firewalls. We examine a late-time scaling limit in JT gravity in which one can resum the expansion in the number of wormholes, and we use this to study the exact distribution of interior slices at times exponential in the entropy. We consider a thermofield double with and without early perturbations on a boundary. These perturbations can appear on interior slices as dangerous high energy shockwaves. For exponentially late times, wormholes tend to teleport the particles created by perturbations and render the interior more dangerous. In states with many perturbations separated by large times, the probability of a safe interior is exponentially small, even though these would be safe without wormholes. With perturbation, even in the safest state we conceive, the odds of encountering a shock are fifty-fifty. One interpretation of the phenomenon is that wormholes can change time-ordered contours into effective out-of-time-ordered folds, making shockwaves appear in unexpected places.
\end{abstract}

\pagebreak
\setcounter{page}{1}
\tableofcontents

\newpage

\section{Introduction}
\label{sec:intro}

The AdS/CFT correspondence~\cite{Maldacena:1997re} has proven to be a very useful tool to understand the rules of quantum gravity. In particular, much recent progress on quantum gravity stems from attempting to find a bulk explanation for exotic physics at late times in the (often chaotic~\cite{Shenker:2013pqa,Cotler:2016fpe}) dual quantum system. Examples include the non-decaying (and highly oscillatory) behavior or late-time correlators of an eternal black hole~\cite{Maldacena:2001kr} and unitarity of black hole evaporation embodied by the Page curve~\cite{Page:1993df}. Both phenomena are explained on the AdS side by (spacetime) wormholes~\cite{Saad:2018bqo,Saad:2019pqd,Saad:2019lba,Blommaert:2019hjr,Iliesiu:2021ari, Kruthoff:2022voq,Blommaert:2020seb,Penington:2019kki,Almheiri:2019qdq}. This is a category of questions which was easier to answer in the CFT than in AdS, which taught us the importance of wormholes.

However, other physically interesting questions about black holes in AdS have proven to be difficult to translate into simple questions in CFT. In particular, this includes the experiences of an observer falling into or residing in a black hole interior. Perhaps those questions are easier to tackle from the bulk point of view, relying on previous lessons on quantum gravity due to AdS/CFT. In this spirit, Stanford and Yang recently asked~\cite{Stanford:2022fdt} if spacetime wormholes have an impact on the firewall question~\cite{Almheiri:2012rt,Almheiri:2013hfa,Marolf:2013dba}.

These authors considered the geometry dual~\cite{Maldacena:2013xja} to the (time evolved) thermofield double (TFD), and found that a single wormhole has the potential to shorten (or lengthen) the Einstein--Rosen bridge (ERB), as will be shown in~\eqref{eq:AT-intro}. This can even turn expanding time slices into contracting slices (the dual of the TFD at negative times). At times of order the Heisenberg time~\cite{Haake:1315494} (or inverse level spacing)%
\footnote{This Heisenberg time is equal to the plateau time~\cite{Cotler:2016fpe}.}
\begin{equation}
  T_H = 2\pi\, e^{S(E)},
\end{equation}
the correction due to the single wormhole amplitude to the total probability of finding an expanding or contracting slice, was found to compete with the leading disk contribution. This raises two immediate questions:
\begin{enumerate}
\item If the one-wormhole contribution competes with the no-wormhole one, then one should also consider contributions from any number of wormholes (and potentially non-perturbative corrections to that series). What is the final distribution of dual semiclassical geometries for $T \sim T_H$?
\item Neither positive nor negative time slices of the TFD have firewalls in the setup that we will study (and with our definition of firewalls), which we will detail below. Potentially more dangerous states have matter perturbations on the boundary in the preparation of the state. Including wormholes, and given any initial state with matter perturbations, what are the odds of encountering a firewall at exponentially late times $T \sim T_H$?
\end{enumerate}

Like Stanford and Yang, we study these questions in 2d Jackiw--Teitelboim (JT) dilaton gravity~\cite{jackiw1985lower,teitelboim1983gravitation,Engelsoy:2016xyb,Jensen:2016pah,Maldacena:2016upp,Mertens:2022irh}. As this theory does not include matter, there is not much that could endanger an observer crossing the horizon in the TFD. On the other hand, early matter perturbations in the preparation of the state grow up to be high energy shockwaves in the bulk~\cite{Shenker:2013pqa,Shenker:2013yza}. Collision with shocks with exponentially high energy \emph{are} harmful. We include matter perturbations in our setup, and find that this can have major effects, deviating significantly from the result of Stanford and Yang~\cite{Stanford:2022fdt}.
The questions of black holes at exponentially late times were also analyzed by Susskind using complexity geometry~\cite{Susskind:2020wwe}.

We use the criterion that \textbf{bulk slices are dangerous if there is at least one strong shockwave in the slice}. By strong, we mean that the shock significantly affects the length of the dual slice. This is what we will mean by the (loaded) word ``firewall,'' nothing more.%
\footnote{There are certainly other sensible definitions of firewalls and ``experiments'' to diagnose them, which  would be interesting to investigate. We believe that our techniques could be modified without too much effort to different setups. An important issue which we do not explicitly address in this paper is the relation of our analyses to the property of the horizon of a collapse-formed single-sided black hole. The construction of~\cite{Nomura:2018kia,Nomura:2019qps,Nomura:2019dlz,Langhoff:2020jqa,Nomura:2020ska,Murdia:2022giv} suggests that the effective two-sided black hole that emerges from such a black hole (after it stabilizes) is the one with $T \ll T_H$ with essentially no perturbation, implying that an infalling observer would not encounter a firewall. We leave a closer investigation of this to the future.}
We will see that wormholes, asides from changing the bare length of the slice~\cite{Stanford:2022fdt}, can also teleport the matter perturbations, resulting in shockwaves in unexpected places. We compute the resulting odds of having a firewall in the bulk slice.

On a technical level, we obtain all-genus results in a relatively straightforward way for two reasons. Firstly, unlike~\cite{Stanford:2022fdt} we define our probability ``operationally''; see~\eqref{eq:defdistribution}. By this we mean they represent density matrices in which you compute expectation values of operators. After all, if we want to imagine that the slice is related to the experiences of an observer, then clearly some type of measurement on the slice must be imagined. This removes the mapping class group subtleties of~\cite{Stanford:2022fdt}. Secondly, we will work in the ``tau-scaling'' limit~\cite{Okuyama:2020ncd,Okuyama:2018gfr,Saad:2022kfe,Blommaert:2022lbh}
\begin{equation}
  T\to \infty\,,\quad T_H\to\infty \,,\quad T/T_H\text{ fixed,}
\label{eq:tauscaling}
\end{equation}
for all times $T$ in the problem. In this regime, we can use exact results from random matrix theory~\cite{Saad:2019lba,mehta2004random} to account for summing over any number of wormholes in a simple manner. We will furthermore consider a regime where the Schwarzian can be treated semi-classically, as in~\cite{Stanford:2022fdt}. In combination, this gives a geometric (semi-classical) interpretation of the full amplitudes.

While our setup evades the mapping class group subtleties of~\cite{Stanford:2022fdt}, it is not clear which setup most accurately models an infalling observer. Our setup has the advantage of being more closely related to a measurement that can be performed on boundaries, and allows on a technical level an incorporation of all-genus effects. The setup of~\cite{Stanford:2022fdt} has the advantage that it has no formal divergences we encounter in section~\ref{sect:gray}. To reach a more definite conclusion for the experience of an infalling observer, we would need a more realistic model of the observer.

\subsection{Summary and structure}

We now summarize the main results of our analysis.

In \textbf{section~\ref{sec:length}}, we study time evolution of the unperturbed TFD \cite{Stanford:2022fdt}. The classical dual geometry (ignoring wormholes) is the time $T$ slice of the two-sided black hole. Quantum mechanically (including wormholes), we instead find that there is a nonzero probability for the geometry to be the slice of the two-sided black hole with \textbf{effective age} $T_\text{eff} \neq 2T$. Schematically, the amplitude decomposes as%
\footnote{We have only depicted the preparation of the ket, which will be glued onto an identical geometry preparing the bra at the red (interior) slice. The Euclidean preparation region is implicit. The factor of $2$ between $T$ and $T_\text{eff}$ arises because $T$ reflects two-sided time evolution while $T_\text{eff}$ can be thought of as one-sided evolution. The latter will be more natural for later sections.}
\begin{equation}
  \begin{tikzpicture}[baseline={([yshift=-.5ex]current bounding box.center)}, scale=0.7]
    \pgftext{\includegraphics[scale=1]{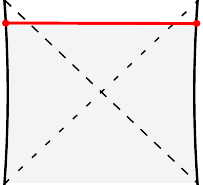}} at (0,0);
    \draw (-2.2,1.2) node {$\color{red}T$};
  \end{tikzpicture}\,\, +\text{ wormholes } 
  \,=\, \mathcal{A}(2T)
  \,=\, \int_{-\infty}^{+\infty}\d T_\text{eff}\,\, \mathcal{F}(T_\text{eff}\rvert 2T)
  \begin{tikzpicture}[baseline={([yshift=-.5ex]current bounding box.center)}, scale=0.7]
    \pgftext{\includegraphics[scale=1]{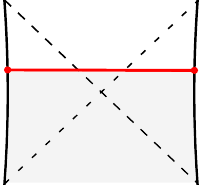}} at (0,0);
    \draw (-2.3,0.4) node {$\color{red}\frac{T_\text{eff}}{2}$};
  \end{tikzpicture}\,\,.
\label{eq:AT-intro}
\end{equation}
The ``conversion factor'' $\mathcal{F}(T_\text{eff}\rvert 2T)$ captures the physical idea that wormholes can change the effective age of the ERB~\cite{Saad:2019pqd,Stanford:2022fdt}. This amplitude has support also for negative effective times $T_\text{eff}<0$, which slice through the white hole interior. One can ask~\cite{Stanford:2022fdt} about the chances of finding a black hole (expanding slice) or white hole (contracting slice) as a function of $T$. This is the integrated conditional probability
\begin{equation}
  P_\text{exp}(2T) = \int_0^{+\infty}\d T_\text{eff}\, \mathcal{F}(T_\text{eff}\rvert 2T)\,,\qquad P_\text{cont}(2T) = \int_{-\infty}^{0}\d T_\text{eff}\, \mathcal{F}(T_\text{eff}\rvert 2T)\,.
\end{equation}
A version of this computation, including the correction from one wormhole, was carried out in~\cite{Stanford:2022fdt} (see also~\cite{Zolfi:2024ldx}).
We do non-perturbative calculation in the tau-scaling limit, which sums the contribution of any number of wormholes, and find that before the Heisenberg time $2T < T_H = 2\pi\, e^{S(E)}$
\begin{equation}
  P_\text{exp}(2T) = 1-\frac{2T}{T_H}+\frac{1}{2}\frac{(2T)^2}{T_H^2}\,,\qquad P_\text{cont}(2T) = \frac{2T}{T_H}-\frac{1}{2}\frac{(2T)^2}{T_H^2}\,.
\end{equation}
Moreover, after a Heisenberg time $2T > T_H$ one finds exactly the \textbf{gray hole} result anticipated in~\cite{Stanford:2022fdt,Susskind:2015toa}, where expanding and contracting slices are equally likely $P_\text{exp}(2T) = P_\text{cont}(2T) = 1/2$. Combining these piecewise functions leads to the following behavior:
\begin{equation}
  \begin{tikzpicture}[baseline={([yshift=-.5ex]current bounding box.center)}, scale=0.7]
    \pgftext{\includegraphics[scale=1]{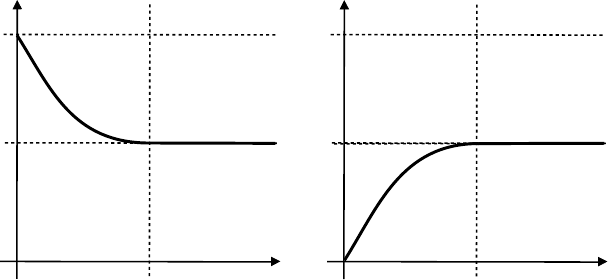}} at (0,0);
    \draw (0,-0.1) node {$1/2$};
    \draw (4.1,-0.7) node {$P_\text{cont}(2T)$};
    \draw (-1.4,-0.7) node {$P_\text{exp}(2T)$};
    \draw (0,1.8) node {$1$};
    \draw (-2.6,-2.8) node {$T_H$};
    \draw (2.9,-2.8) node {$T_H$};
  \end{tikzpicture}\label{1.6intro}
\end{equation}
Given the close mathematical relation with the famous plateau in the spectral form factor~\cite{Cotler:2016fpe}, we call the $2T>T_H$ behavior in~\eqref{1.6intro} the \textbf{firewall probability plateau}.

In \textbf{section~\ref{sec:one}}, we will study the TFD with one thermal perturbation at early times $t=-T_w$ on the left boundary at Heisenberg time scales, $T, T_w \sim T_H$. In JT gravity, neither expanding nor contracting slices of the TFD are dangerous. Potential danger arises from high energy particles that cross the slices. In our setup, such particles may arise only from perturbations that were added in the preparation of the state.%
\footnote{We will consider JT gravity with probe matter, not a (dynamical) matter QFT coupled to JT gravity. It would be very interesting to study also the latter. Unfortunately, however, studying dynamical matter on wormholes in JT is notoriously challenging~\cite{Saad:2019lba,Moitra:2021uiv} (but see~\cite{Jafferis:2022wez,Belaey:2023jtr}). We will not attempt this here, but comment more on this in the discussion section~\ref{sec:concl}.}
We find that wormholes may teleport the perturbations far to the past or future, changing the \textbf{effective perturbation time} $T_{w\,\text{eff}}\neq T_w$, and therefore potentially rendering naively safe encounters into dangerous ones (and vice versa).

We will show that the amplitude at time $T$ decomposes (in our regime of approximation) as
\begin{equation}
    \begin{tikzpicture}[baseline={([yshift=-.5ex]current bounding box.center)}, scale=0.7]
        \pgftext{\includegraphics[scale=1]{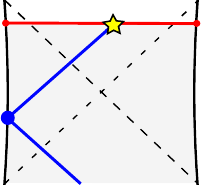}} at (0,0);
        \draw (-2.2,1.2) node {$\color{red}T$};
        \draw (-2.2,-0.5) node {$\color{blue}T_w$};
    \end{tikzpicture}\quad+\text{ wormholes }=\int_{-\infty}^{+\infty}\d T_\text{eff}\,\d T_{w\,\text{eff}}\, \mathcal{G}(T_\text{eff},T_{w\,\text{eff}}\rvert T,T_w)\,
    \begin{tikzpicture}[baseline={([yshift=-.5ex]current bounding box.center)}, scale=0.7]
        \pgftext{\includegraphics[scale=1]{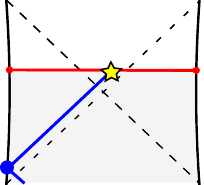}} at (0,0);
        \draw (-2.3,0.4) node {$\color{red}T_\text{eff}$};
        \draw (-2.5,-1.2) node {$\color{blue}T_{w\,\text{eff}}$};
    \end{tikzpicture}
\label{eq:G-conversion}
\end{equation}
with some conversion factor $\mathcal{G}(T_\text{eff},T_{w\,\text{eff}}\rvert T,T_w)$. Introducing $T_1=T-T_w$ and $T_2=T+T_w$, we find that the conversion factor factorizes to a product of conversion factors appearing in~\eqref{eq:AT-intro}:\ $\mathcal{G}(T_\text{eff},T_{w\,\text{eff}}\rvert T,T_w) \sim \mathcal{F}(T_{1\,\text{eff}}\rvert T_1)\mathcal{F}(T_{2\,\text{eff}}\rvert T_2)$. This has support for all signs of $T_{1\,\text{eff}}, T_{2\,\text{eff}}$, and therefore wormholes in this observable offer the possibility to essentially \textbf{turn time-ordered contours into effective out-of-time ordered folds}, and vice versa. Here we are referring to the time-folds that arise in the boundary dual when preparing the state at the late time slice of interest, starting from the $t=0$ TFD.%
\footnote{See equation~(4.22) and Figure~6 in~\cite{Stanford:2014jda} or our equation~\eqref{eq:timefolds}.}
Adhering to the notation of~\cite{Stanford:2014jda}, we can indeed write~\eqref{eq:G-conversion} schematically as
\begin{equation}
    \begin{tikzpicture}[baseline={([yshift=-.5ex]current bounding box.center)}, scale=0.7]
        \pgftext{\includegraphics[scale=1]{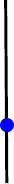}} at (0,0);
        \draw (-0.6,0.6) node {$T_2$};
        \draw (-0.6,-1.2) node {$T_1$};
    \end{tikzpicture}\quad+\text{ wormholes }=\int_{-\infty}^{+\infty}\d T_{1\,\text{eff}}\,\mathcal{F}(T_{1\,\text{eff}}\rvert T_1)\int_{-\infty}^{+\infty}\d T_{2\,\text{eff}}\,\mathcal{F}(T_{2\,\text{eff}}\rvert T_2)\quad
    \begin{tikzpicture}[baseline={([yshift=-.5ex]current bounding box.center)}, scale=0.7]
        \pgftext{\includegraphics[scale=1]{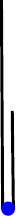}} at (0,0);
        \draw (0.9,-0.9) node {$T_{1\,\text{eff}}$};
        \draw (-0.95,0.2) node {$T_{2\,\text{eff}}$};
    \end{tikzpicture}\,\,.
\label{1.4}
\end{equation}
We use these time-folds as shorthand notation for classical slices being probed in the dual geometry.

We consider the interior slice to be \textbf{dangerous} if and only if there is a strong shockwave, meaning that a shock severely backreacts on the geometry of the effective slice.%
\footnote{Technically, in our JT gravity setup the discriminator is whether or not the length of the slice significantly changes due to shocks.}
This happens \textbf{if there is at least one switchback in the effective time-fold} preparing the state~\cite{Stanford:2014jda}. At very late times, all signs of $T_{i\,\text{eff}}$ are equally likely. The options of $(T_{1\,\text{eff}}, T_{2\,\text{eff}})$ being $(++)$ and $(--)$ have no switchbacks. The options $(+-)$ and $(-+)$ have a switchback and are dangerous. Therefore, \textbf{the odds of encountering a firewall at very late times are even} in this setup
\begin{equation}
  \boxed{P_\text{safe}(T)=P_\text{danger}(T)=\frac{1}{2}}\,,\quad T>T_H+T_w\,.
\label{eq:intro:shock-half}
\end{equation}
In \textbf{appendix~\ref{app:avoided}}, we show that another effect of wormholes, namely the possibility that the perturbation is carried away by a wormhole (thus avoiding a collision in the interior), vanishes in the tau-scaling limit.

In \textbf{section~\ref{sec:multiple}}, we consider more general states obtained by perturbing the TFD on the left boundary at multiple times $t=-T_{w_i}$ ($i=1,\ldots,n$), where $T_{w_1} < \ldots < T_{w_n}$. We will consider several different configurations. The worst case scenario is one where $n$ perturbations are each separated by more than a Heisenberg time, in which case \textbf{dangerous interiors are exponentially likely} in $n$
\begin{equation}
  \boxed{P_\text{safe}(T)=\frac{1}{2^n}}\,\,\, \overset{n\gg 1}{\longrightarrow} \,\, 0\,,\quad T>T_{w_n}+T_H\,.
\label{eq:intro-firewall}
\end{equation}
This is simple to see using the generalization of~\eqref{1.4}. The effective time-fold consists of $n+1$ segments
\begin{equation}
    \begin{tikzpicture}[baseline={([yshift=-.5ex]current bounding box.center)}, scale=0.7]
        \pgftext{\includegraphics[scale=1]{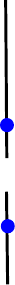}} at (0,0);
        \draw (-0.8,1.5) node {$T_{n+1}$};
        \draw (0,-0.58) node {$\dots$};
        \draw (-0.6,-2) node {$T_1$};
    \end{tikzpicture}\quad+\text{ wormholes }=\prod_{i=1}^{n+1}\int_{-\infty}^{+\infty}\!\d T_{i\,\text{eff}}\,\mathcal{F}(T_{i\,\text{eff}}\rvert T_i)\quad
    \begin{tikzpicture}[baseline={([yshift=-.5ex]current bounding box.center)}, scale=0.7]
        \pgftext{\includegraphics[scale=1]{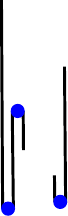}} at (0,0);
        \draw (1.3,-0.5) node {$T_{1\,\text{eff}}$};
        \draw (-1.6,0.2) node {$T_{n+1\,\text{eff}}$};
        \draw (0.1,-0.9) node {$\dots$};
    \end{tikzpicture}\,\,.
\label{eq:manyfold-intro}
\end{equation}
For sufficiently late times, the sign of the times $T_{i\,\text{eff}}$ of each segment is random. The only safe scenarios are when there are no switchbacks, so all signs of $T_{i\,\text{eff}}$ must match. This is exponentially unlikely in $n$. We consider other states, but at late times none of them are safer than the single-particle setup \eqref{eq:intro:shock-half}, which is realized if the $n$ perturbations are separated by times much less than the scrambling time $T_S$.

We stress again that what we are computing here is the probability that there is at least one dangerous shock on the interior slice in pure JT gravity. As shown in Figure~4 in~\cite{Shenker:2013yza}, on a multiple shockwave geometry in higher dimensions, only the outermost shockwave determines the experience of an infalling observer. In this case, we would conclude that even with multiple shock waves, the probability of firewall at $T\to\infty$ would still be $P=1/2$.\footnote{We thank Douglas Stanford for pointing this out.} In pure JT gravity, however, there is no spacelike singularity, so the probability of an infalling observer not being hit by a shock is~\eqref{eq:intro-firewall}.

Finally, in \textbf{section~\ref{sec:concl}} we reemphasize that these results are obtained by summing a perturbatively convergent series (in genus) of wormhole amplitudes, following~\cite{Saad:2022kfe,Blommaert:2022lbh,Weber:2022sov}, and identify the corresponding Lorentzian wormhole geometries~\cite{Blommaert:2023vbz,Usatyuk:2022afj}. These explicitly realize the notion of effective times. We will also identify several shortcomings of our work, making concrete proposals for how to improve on it.

\subsection*{Relation to other work}
Similar question have been independently investigated at the same time via related techniques by Iliesiu, Levine, Lin, Maxfield and Mezei~\cite{wipluca}, with whom we have coordinated submissions.

\section{Thermofield double at exponentially late times}
\label{sec:length}

In this section, we study the effects of wormholes on the time evolution of the unperturbed TFD state in JT gravity. In particular, we will ask about the distribution of the length and bulk spatial slices as a function of boundary time $T$. From here on, we will always picture purely Euclidean spacetimes, with the appropriate Lorentzian continuation implied by the boundary conditions.

\subsection{Setup}
\label{sect:2.1}

Suppose one prepares the TFD at $t=0$, and time evolve this state on both sides during boundary time $T$. We are interested in the bulk interpretation of the resulting state. We imagine this is diagnosed by performing simple two-sided measurements, such as computing two-sided two-point functions.%
\footnote{We could more generally consider two-sided higher point functions (with operators inserted at the same boundary time) which in the tau-scaling limit are computed using the same distribution $\tilde{\mathcal{F}}(\ell\rvert 2T)$ in~\eqref{eq:Al_def}. Off-diagonal contributions in the $\ell$-basis are then suppressed by powers of $e^{\S}$, since without exponentially large times weighting the boundary segments between operator insertions, there is nothing canceling such suppression.}
In JT gravity those are computed by summing over wormholes~\cite{Saad:2019pqd,Blommaert:2019hjr,Blommaert:2020seb,Iliesiu:2021ari}%
\footnote{The argument is $2T$ because we are evolving both the left and right sides of the TFD by $T$.}
\begin{equation}
  G_{\Delta\,\text{nonpert}}(2T) =
    \begin{tikzpicture}[baseline={([yshift=-.5ex]current bounding box.center)}, scale=0.7]
        \pgftext{\includegraphics[scale=1]{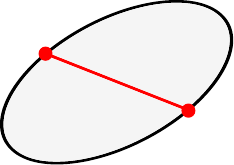}} at (0,0);
        \draw (-2,0.7) node {$\color{red}\mathcal{O}_\Delta$};
        \draw (2,-0.7) node {$\color{red}\mathcal{O}_\Delta$};
        \draw (2.6,0.6) node {$\b_1$};
        \draw (-2.65,-0.7) node {$\b_2$};
    \end{tikzpicture}
  \quad+\quad
    \begin{tikzpicture}[baseline={([yshift=-.5ex]current bounding box.center)}, scale=0.7]
        \pgftext{\includegraphics[scale=1]{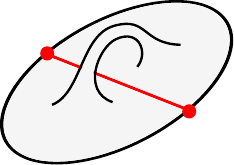}} at (0,0);
        \draw (-2,0.7) node {$\color{red}\mathcal{O}_\Delta$};
        \draw (2,-0.7) node {$\color{red}\mathcal{O}_\Delta$};
        \draw (2.6,0.6) node {$\b_1$};
        \draw (-2.65,-0.7) node {$\b_2$};
    \end{tikzpicture}
  + \quad\dots 
\label{eq:euclwh}
\end{equation}
with analytic continuation of the boundary conditions to Lorentzian times
\begin{equation}
  \beta_1=\frac{\beta}{2}+2\i T\,,\quad \beta_2=\frac{\beta}{2}-2\i T\,.
\end{equation}
Here, as announced, we have depicted the Euclidean AdS$_2$ wormhole geometries.
These path integrals can be computed exactly and the full amplitude decomposes as~\cite{Saad:2019pqd,Blommaert:2019hjr,Blommaert:2020seb,Iliesiu:2021ari}
\begin{equation}
  G_{\Delta\,\text{nonpert}}(2T) = \int_{-\infty}^{+\infty} \d\ell\,\tilde{\mathcal{F}}(\ell\rvert 2T)\, G_{\Delta\,\text{class}} (\ell)\,,
\label{eq:Al_def}
\end{equation}
where for boundary operators of conformal weight $\Delta$ (dual to a bulk particle of mass $\Delta$)~\cite{Saad:2019pqd,Blommaert:2018oro,Yang:2018gdb}
\begin{equation}
  G_{\Delta\,\text{class}}(\ell) = e^{-\Delta \ell}\,.
\end{equation}
Here $\ell$ is the geodesic distance covered by the particle. In JT gravity this decomposition \eqref{eq:Al_def} is exact.

The quantity of our interest is a slightly modified version of the distribution $\tilde{\mathcal{F}}(\ell\rvert 2T)$
\begin{equation}
  G_{\Delta\,\text{nonpert}}(2T) = \int_{-\infty}^{+\infty} \d T_\text{eff}\,\mathcal{F}(T_\text{eff}\rvert 2T)\, G_{\Delta\,\text{class}}(T_\text{eff})\,.
\label{eq:defdistribution}
\end{equation}
$T_\text{eff}$ is defined to be the length of the spatial slice in the classical two-sided black hole when {\it one of} the boundaries is evolved by $T_\text{eff}$, or equivalently both boundaries are evolved by $T_\text{eff}/2$.
(This definition is more convenient in later sections.)
To be precise, in JT gravity the time-evolving TFD spacetime has the metric and dilaton%
\footnote{Even though we use the terminology TFD we will almost exclusively study a microcanonical ensemble with fixed energy $E$. The Lorentzian spacetime for a fixed energy is also smooth~\cite{Dong:2022ilf} and classically $E = \pi^2/\beta^2$.}
\begin{equation}
  \begin{tikzpicture}[baseline={([yshift=-.5ex]current bounding box.center)}, scale=0.7]
        \pgftext{\includegraphics[scale=1]{fire4.pdf}} at (0,0);
        \draw (-2.3,0.4) node {$\color{red}X$};
  \end{tikzpicture}
  \qquad
  \d s^2 = \frac{-\d X^2 + \d\sigma^2}{\sin(\sigma)^2}\,,\quad \Phi=E^{1/2}\,\frac{\cos(X)}{\sin(\sigma)}\,.
\label{eq:geom}
\end{equation}
In these conformal coordinates, the location of the boundary changes as a function of boundary time $t$ as (see for instance~\cite{Maldacena:2016upp} for more details)
\begin{equation}
  \tan(X/2) = \tanh(E^{1/2} t)\,,\quad \sigma = \pi - \varepsilon\, \frac{\d X}{\d t}\,.
\end{equation}
The geodesic between boundary points at $t = T_\text{eff}/2$ is at constant $X$ and has (renormalized) length
\begin{equation}
  \boxed{e^{\ell/2}=\frac{1}{E^{1/2}}\cosh(E^{1/2}\,T_\text{eff})}\,.
\label{eq:ell}
\end{equation}

The correct interpretation of~\eqref{eq:defdistribution} is that, after including wormhole corrections, the actual Lorentzian spacetime one is probing is a distribution of slices of the ordinary TFD with effective ages $T_\text{eff}$ different from $2T$. One way to see this is to notice that $G_{\Delta\,\text{class}}(T_\text{eff})$ is the classical two-point function in such a spacetime. We will also see in section~\ref{sect:semiwave} that this indeed follows from a semiclassical analysis of the exact JT amplitudes, combined with a geometric interpretation of all integration parameters~\cite{Yang:2018gdb,Stanford:2021bhl}. Finally, this interpretation also results because the purely Lorentzian analogues of the Euclidean wormhole geometries sketched in~\eqref{eq:euclwh} are actually known~\cite{Blommaert:2023vbz}, where the probed Lorentzian slices are indeed $T_\text{eff}$ slices of the TFD. We will discuss those Lorentzian wormhole geometries in more detail in section~\ref{sect:disclor}.

Before we compute and analyze $\mathcal{F}(T_\text{eff}\rvert 2T)$, let us make two comments about our setup:
\begin{enumerate}
\item
Our conversion factor $\mathcal{F}(T_\text{eff}\rvert 2T)$, even though very similar in spirit, is mathematically different from the quantity $\mathcal{F}_{\text{SY}}(T_\text{eff}\rvert 2T)$ computed by Stanford and Yang~\cite{Stanford:2022fdt}, who define
\begin{equation}
  Z(2T) = \int_{-\infty}^{+\infty}\d \ell\, \tilde{\mathcal{F}}_{\text{SY}}(\ell\rvert 2T)\, Z(\ell)\,.
\end{equation}
They decompose the partition function \emph{without} operator insertions in the length basis. For famous mathematical reasons having to do with the mapping class group, their calculation is actually a more complicated problem in JT gravity than ours. This topic is very didactically documented, starting with~\cite{mirzakhani2007simple,Dijkgraaf:2018vnm,Stanford:2019vob}. How this plays together with a calculation of the two-point function in JT gravity was explained in~\cite{Blommaert:2019hjr,Saad:2019pqd,Blommaert:2020seb}, and reviewed nicely in~\cite{Iliesiu:2021ari}. We will not review it again here.
\item
Our setup is at least as well motivated as that of Stanford and Yang. Indeed, without measurement there is not much physical relevance to amplitudes. The density matrix that is actually appearing in measurements is the amplitude we study in~\eqref{eq:defdistribution}. We do not claim that the setup of~\cite{Stanford:2022fdt} is not reasonable to study. As an additional a posteriori justification of our method, we note that our final answer~\eqref{1.6intro} is physically sensible. We emphasize that we can not claim that infalling observers are modeled well by the simple two-sided measurements we consider here, even though this is the ultimate goal we must have in mind. See also section~\ref{sec:concl}.
\end{enumerate}
In the remainder of this section, we will compute and analyze $\mathcal{F}(T_\text{eff}\rvert 2T)$.

\subsection{Exact answer from the gravitational path integral}

We start with an exact expression for the amplitude of the two-point function in JT gravity~\cite{Saad:2019pqd,Blommaert:2019hjr,Blommaert:2020seb,Iliesiu:2021ari}
\begin{align}
  &G_{\Delta\,\text{nonpert}}(2T) = \int_{-\infty}^{+\infty} \d\ell\,\tilde{\mathcal{F}}(\ell\rvert 2T)\, e^{-\Delta \ell}
\label{eq:2point-JT}\\
  &\quad =\frac{1}{Z(\beta)} \int_0^\infty\! \d E_1\,e^{-(\beta/2+2\i T)E_1} \int_0^\infty\! \d E_2\,e^{-(\beta/2-2\i T)E_2}\, \rho(E_1,E_2)\,e^{-\S}\! \int_{-\infty}^{+\infty}\!\d\ell\, \psi_{E_1}(\ell)\psi_{E_2}(\ell)\,e^{-\Delta \ell}\nonumber
\end{align}
with orthonormal wavefunctions
\begin{equation}
    \psi_E(\ell)=4 K_{2\i E^{1/2}}\big(e^{-\ell/2}\big)\,,\quad \int_{-\infty}^{+\infty}\d\ell\, \psi_{E_1}(\ell)\psi_{E_2}(\ell)=\frac{\delta(E_1-E_2)}{\rho_0(E)}\,,\quad \rho_0(E)=\frac{\sinh(2\pi E^{1/2})}{4\pi^2}\,,
\label{eq:bessels}
\end{equation}
where $\S$ is the extremal entropy \cite{Saad:2019lba}.
The integration measure $\rho(E_1,E_2)$ or ``spectral correlation'' is
\begin{equation}
  \rho(E_1,E_2)=\rho(E_1)\rho(E_2)+\rho(E_1,E_2)_\text{conn}\,,
\end{equation}
with \cite{Saad:2019lba}
\begin{equation}
  \rho(E) = e^{\S}\rho_0(E)+\sum_{g=1}^\infty e^{(1-2g)\S}\int_0^\infty \d b\, \rho_\text{trumpet}(E,b)\,V_{g,1}(b)+\text{ non-perturbative,}
\end{equation}
and
\begin{align}
  &\rho(E_1,E_2)_\text{conn}\nonumber\\&=\sum_{g=0}^\infty e^{-2g\S}\int_0^\infty \d b_1\, \rho_\text{trumpet}(E_1,b_1)\int_0^\infty \d b_2\, \rho_\text{trumpet}(E_2,b_2)\,V_{g,2}(b_1,b_2)+\text{ non-perturbative.}
\label{eq:rho_conn}
\end{align}
Here, the ``trumpet'' density of states is~\cite{Saad:2019lba}
\begin{equation}
  \rho_\text{trumpet}(E,b)=\frac{\cos(b E^{1/2})}{2\pi E^{1/2}}\,,
\end{equation}
and $V_{g,n}(b_1\dots b_n)$ are Weil-Peterson volumes of the moduli spaces of Riemann surfaces~\cite{mirzakhani2007simple,Dijkgraaf:2018vnm,Stanford:2019vob}.

The genus $g$ expansions are completed non-perturbatively by a matrix integral answer~\cite{Saad:2019lba}. However, if in addition to the large $\S$ limit, we also take the late-time limit, namely the tau-scaling limit,~\eqref{eq:tauscaling}
\begin{equation}
  T\to\infty\,,\quad e^{\S}\to\infty\,,\quad Te^{-\S}\text{ fixed,}
\end{equation}
then this story simplifies significantly. In fact, the genus expansion reproduces already~\eqref{eq:sinekernel}~\cite{Stanford:2022fdt,Blommaert:2022lbh}; \emph{no} non-perturbative completion is needed! This will be the regime of interest in this paper.

To further elaborate, let us introduce the parametrization
\begin{equation}
  E_1=E+\frac{\omega}{2}\,,\quad E_2=E-\frac{\omega}{2}\, ,
\end{equation}
so that the Boltzmann weights in~\eqref{eq:2point-JT} become $e^{-\beta E}e^{-2\i\omega T}$\,. For $T\sim e^{\S}\to\infty$ the Fourier transform (the $\omega$ integral) receives contributions from the least analytic components of $\rho(E_1,E_2)$ as a function of $\omega$. They happen to arise from the tau-scaling limit in the energy domain
\begin{equation}
  \omega\to 0\,,\quad e^{\S}\to\infty\,, \quad \omega\,e^{\S}\text{ fixed.}
\label{eq:tau-scaling_omega}
\end{equation}
This is precisely the regime~\cite{efetov1983supersymmetry} in which correlators in random matrix theory have a universal answer, featuring the so-called sine kernel~\cite{mehta2004random,Haake:1315494}
\begin{equation}
  \rho(E) = e^{\S}\rho_0(E)\,,\quad \rho(E_1,E_2) = \rho(E)^2+\delta(\omega)\rho(E)-\frac{\sin(\pi\rho(E)\omega)^2}{\pi\omega^2}\,,
\label{eq:sinekernel}
\end{equation}
where $\rho(E) = e^{S(E)}$.
Importantly, the argument of the sine function contains a factor of $e^{\S}$ from the density of states. The generalization of \eqref{eq:sinekernel} to multiple energy entries $\rho(E_1\dots E_n)$ is well known~\cite{mehta2004random} and will be implicitly used in sections~\ref{sec:one} and \ref{sec:multiple}.%
\footnote{For detailed equations, see for instance (2.26) in~\cite{Blommaert:2020seb} and section~3.2 in~\cite{Blommaert:2019wfy}.}

We will work at fixed $E$ instead of $\beta$ (by doing an inverse Laplace transform) and consider large black holes $E\gg 1$, which suppresses Schwarzian (quantum) fluctuations. At this stage, we have
\begin{equation}
    \tilde{\mathcal{F}}(\ell\rvert 2T)=\frac{e^{-\S}}{\rho(E)}\int_{-\infty}^{+\infty}\d\omega\,e^{-2\i\omega T}\,\psi_{E+\omega/2}(\ell)\psi_{E-\omega/2}(\ell)\,\rho(E_1,E_2)\,
\label{eq:tilde-F}
\end{equation}
with the spectral correlation evaluated as~\eqref{eq:sinekernel}. Below we will discuss the semiclassical approximation of the wavefunctions $\psi_{E+\omega/2}(\ell) \psi_{E-\omega/2}(\ell)$.%
\footnote{The result we will obtain from this is not qualitatively new. In section~2.2 of~\cite{Stanford:2022fdt}, the authors approximate the Bessel functions by cosines, resulting in contributions such as
\begin{align}
  \psi_{E+\omega/2}(\ell)\psi_{E-\omega/2}(\ell) \sim e^{\i \omega \frac{\ell}{2\sqrt{E}}},
\end{align}
Our approximation is more precise, but more importantly, it prepares us well for the more complicated setup of sections~\ref{sec:one}.}

\subsection{Semiclassical wavefunction and effective time}
\label{sect:semiwave}

We claim that the correct semiclassical approximation of the wavefunctions is%
\footnote{An immediate check is the following computation in the tau-scaling limit
\begin{align}
    \int_{-\log(E)}^{+\infty}\!\d \ell\, \frac{1}{\rho(E)}\,\frac{1}{2\pi}\,\frac{1}{(E-e^{-\ell})^{1/2}}\,\cos\bigg(\frac{\omega}{E^{1/2}}\text{arccosh}\Big(e^{\ell/2}E^{1/2}\Big)\bigg)\,e^{-\Delta\ell}&= \int_{-\log(E)}^{+\infty}\d \ell\, \frac{1}{\rho(E)}\,\frac{1}{2\pi}\,\frac{1}{(E-e^{-\ell})^{1/2}}\,\,e^{-\Delta\ell}\nonumber\\&=\frac{1}{2\pi}\frac{1}{\rho(E)}\frac{\Gamma(\Delta)^2}{\Gamma(2\Delta)}e^{-(\Delta-1/2)(-\log(4 E))}\, ,
\end{align}
which matches the semiclassical approximation that one gets by directly taking $E\gg 1$ and $\omega\ll 1$ in the Gamma functions that arise~\cite{Blommaert:2018oro,Yang:2018gdb,Iliesiu:2019xuh} when we compute
\begin{equation}
    \int_{-\infty}^{+\infty}\!\d \ell\,\psi_{E_1}(\ell)\psi_{E_2}(\ell)\,e^{-\Delta\ell}
\end{equation}
using the exact wavefunctions~\eqref{eq:bessels}. In the second step we used the fact that the exponential suppression $e^{-\Delta \ell}$ destroys contributions from lengths of order $e^{\S}$ such that the cosine evaluates to one.}
\begin{equation}
    \psi_{E+\omega/2}(\ell)\psi_{E-\omega/2}(\ell)=\Theta(\ell+\log(E))\,\frac{1}{\rho_0(E)}\,\frac{1}{2\pi}\,\frac{1}{(E-e^{-\ell})^{1/2}}\,\cos\bigg(\frac{\omega}{E^{1/2}}\text{arccosh}\Big(2e^{\ell/2}E^{1/2}\Big)\bigg)\,.
\label{eq:psi-psi}
\end{equation}
Let us derive this. The integral representations of the Bessel functions gives
\begin{align}
  &\psi_{E+\omega/2}(\ell)\psi_{E-\omega/2}(\ell)
\label{eq:integral-rep}\\
  &\quad =\int_{-\infty}^{+\infty}\d b_1\,\d b_2\exp\bigg( \i (b_1+b_2) E^{1/2}+\i (b_1-b_2)\frac{\omega}{4 E^{1/2}}-2 e^{-\ell/2}\cosh(\frac{b_1+b_2}{4})\cosh(\frac{b_1-b_2}{4}) \bigg).
\nonumber
\end{align}
Introducing ``angular'' variables $\a_i$ (the geometrical meaning of which we will discuss shortly)
\begin{equation}
  2\pi+\i b_i=\a_i
\label{eq:alpha_i}
\end{equation}
and furthermore relabeling these integration variables as
\begin{equation}
  \alpha_1=\pi+2\i E^{1/2}(T_\text{eff}+\Delta T_\text{eff})\,,\quad \alpha_2=\pi+2\i E^{1/2}(-T_\text{eff}+\Delta T_\text{eff}),
\end{equation}
the integrand becomes
\begin{equation}
    \exp\bigg( -2\pi E^{1/2}+\i \omega T_\text{eff}+4\i E \Delta T_\text{eff}-2\i e^{-\ell/2} \cosh (T_\text{eff} E^{1/2})\sinh(\Delta T_\text{eff}E^{1/2})\bigg)\,.
\label{eq:integrand}
\end{equation}

Without wormhole corrections, represented by the second and third terms of $\rho(E_1,E_2)$ in~\eqref{eq:sinekernel}, one could at this point already do the $\omega$ integral in~\eqref{eq:tilde-F} and find $T_\text{eff}=2T$. This is the classical answer on the disk, where $\alpha_i$ have the interpretation~\cite{Yang:2018gdb,Stanford:2021bhl} of angles on the Euclidean disk
\begin{equation}
  \alpha_i=\frac{2\pi}{\beta}\,\beta_i\,.
\label{eq:alpha_i-disk}
\end{equation}
In our notation, $\beta_1=\b/2+2\i T$ and $\beta_2=\b/2-2\i T$ with $\pi/\beta=E^{1/2}$. Then one indeed finds $T_\text{eff}=2T$ and $\Delta T_\text{eff}=0$.

However, the corrections due to wormholes in the integration kernel, \eqref{eq:sinekernel} in~\eqref{eq:tilde-F}, mean that the $\omega$ integral no longer localizes on $T_\text{eff}=2T$. We find it more convenient to keep the $\omega$ integral in~\eqref{eq:tilde-F} and instead look for saddles of the $T_\text{eff}$ and $\Delta T_\text{eff}$ (or $b_1$ and $b_2$) integrals. The $T_\text{eff}$ and $\Delta T_\text{eff}$ equations of motion are
\begin{equation}
    \omega=2 e^{-\ell/2} E^{1/2}\sinh(T_\text{eff} E^{1/2})\sinh(\Delta T_\text{eff}E^{1/2})\,,\quad e^{-\ell/2} =\frac{2\, E^{1/2}}{\cosh(T_\text{eff} E^{1/2})\cosh(\Delta T_\text{eff} E^{1/2})}\, .
\end{equation}
In the tau-scaling limit~\eqref{eq:tau-scaling_omega} where $\omega\sim e^{-\S}\to 0$, this has the following solutions
\begin{equation}
    \Delta T_\text{eff}=\frac{\omega}{4 E^{3/2}}\to 0\,,\quad T_\text{eff}=\pm \frac{1}{E^{1/2}}\text{arccosh}\Big(2e^{\ell/2}E^{1/2}\Big)\,,
\label{eq:T_Hff}
\end{equation}
resulting in the on-shell actions
\begin{equation}
    \exp \bigg( -2\pi E^{1/2}\pm \i \frac{\omega}{E^{1/2}}\text{arccosh}\Big(2e^{\ell/2}E^{1/2}\Big)\bigg)\,.
\end{equation}
After including the one-loop factors this results in~\eqref{eq:psi-psi}. The Heaviside function arises from the fact that the saddle is only valid for real $T_\text{eff}$.

Since we are at fixed energy, we could do a change of coordinates form $\ell$ to
\begin{equation}
  T_\ell=\frac{1}{E^{1/2}}\text{arccosh}\Big(2e^{\ell/2}E^{1/2}\Big)\,.
\end{equation}
Then the wavefunction-squared would simplify tremendously%
\footnote{As an intermediate check on our approximations here, note that at fixed $\omega$ the expectation value of $T_\ell$ is
\begin{equation}
  \int_{-\infty}^{+\infty}\d T_\ell\,\Theta(T_\ell)\,\frac{1}{\rho_0(E)}\,\frac{1}{\pi}\cos(\omega T_\ell)\,T_\ell=-\frac{1}{\pi}\frac{1}{\rho(E)}\,\text{fp}\bigg(\frac{1}{\omega^2}\bigg)\,,
\end{equation}
where fp is the Hadamard finite part. This matches exactly with equation~(4.5) in~\cite{Iliesiu:2021ari}.}
\begin{equation}
  \d\ell\, \psi_{E+\omega/2}(\ell)\psi_{E-\omega/2}(\ell)=\d T_\ell\,\Theta(T_\ell)\,\frac{1}{\rho_0(E)}\,\frac{1}{\pi}\cos(\omega T_\ell)\,,
\label{eq:distrsemi}
\end{equation}
which we can check is still correctly normalized as in~\eqref{eq:bessels}. We, however, see that it is more convenient to instead write the distribution in terms of $T_\text{eff}=\pm T_\ell$, which takes positive \emph{and} negative values. In conclusion, we find that the conversion factor from $T$ to $T_\text{eff}$, corresponding to that in~\eqref{eq:2point-JT} from $T$ to $\ell$, is (semiclassically)
\begin{equation}
  \mathcal{F}(T_\text{eff}\rvert 2T) = \frac{1}{2\pi \rho(E)^2}\int_{-\infty}^{+\infty}\d\omega\,e^{\i \omega (T_\text{eff}-2T)}\bigg(\rho(E)^2+\delta(\omega)\rho(E)-\frac{\sin(\pi\rho(E)\omega)^2}{\pi\omega^2}\bigg)\,.
\label{eq:F-final}
\end{equation}
The disk contribution (the first term in the parentheses) is the classical answer. The bulk geometry is indeed the expected TFD with both sides evolved by $T$. In other words
\begin{equation}
  \mathcal{F}(T_\text{eff}\rvert 2T)=\delta(T_\text{eff}-2T)+\text{ wormholes.}
\label{cllimit}
\end{equation}

We note that the contribution at fixed $T_\text{eff}$ comes from the saddle
\begin{equation}
  \alpha_1=\pi+2\i E^{1/2}\,T_\text{eff}\,,\quad \alpha_2=\pi-2\i E^{1/2}\,T_\text{eff}\,.
\end{equation}
This is the saddle point of the disk path integral with Euclidean boundary times $\beta_1=\b/2+\i T_\text{eff}$ and $\beta_2=\b/2-\i T_\text{eff}$ on the segments between operator insertions. We will encounter a similar phenomenon in sections~\ref{sec:one} and \ref{sec:multiple}, where physics (the exact amplitude) essentially factorizes into on one hand wormhole physics, which changes the saddles of $\alpha_i$ away from their classical disk answers, and on the other hand particle scattering at fixed $\alpha_i$. The latter reproduces a disk amplitude \emph{as if} we would have put boundary conditions $\beta_{i\,\text{eff}}=\a_i\beta/2\pi$ (with $\alpha_i$ affected by wormholes and integrated over). In this case, this results in the pictorial representation of the amplitude
\begin{equation}
  \begin{tikzpicture}[baseline={([yshift=-.5ex]current bounding box.center)}, scale=0.7]
    \pgftext{\includegraphics[scale=1]{fire3.pdf}} at (0,0);
    \draw (-2.2,1.2) node {$\color{red}T$};
  \end{tikzpicture}\,\, +\text{ wormholes } 
  \,=\, \mathcal{A}(2T)
  \,=\, \int_{-\infty}^{+\infty}\d T_\text{eff}\,\, \mathcal{F}(T_\text{eff}\rvert 2T)
  \begin{tikzpicture}[baseline={([yshift=-.5ex]current bounding box.center)}, scale=0.7]
    \pgftext{\includegraphics[scale=1]{fire4.pdf}} at (0,0);
    \draw (-2.3,0.4) node {$\color{red}\frac{T_\text{eff}}{2}$};
  \end{tikzpicture}
\label{eq:efftime1}
\end{equation}
with $\mathcal{F}(T_\text{eff}\rvert 2T)$ given by~\eqref{eq:F-final}. This is the expression quoted in~\eqref{eq:AT-intro} in the introduction.

We read this equation as showing us the Lorentzian way to interpret Euclidean wormhole calculations. In other words, we interpret geometries with effective age $T_\text{eff}$ as true Lorentzian slices that are being probed when we compute boundary two-point functions in JT gravity, and we want to know the distribution of such dual slices. Let us make several comments regarding this interpretation.
\begin{enumerate}
\item
In the exact answer~\eqref{eq:2point-JT}, $\ell$ is the length of a geodesic in AdS$_2$ to all orders in the genus expansion. A boundary observer measuring this correlator could reach two conclusions. They may take the fact that the answer does not match $e^{-\Delta \ell(2T)}$ as evidence that any notion of classical spacetimes has failed. We believe that this would be a {\it wrong} conclusion, as the exact Euclidean calculation (in the tau-scaling limit) is purely geometric~\cite{Blommaert:2022lbh}. Or they could conclude that they are probing a wavefunction with support on different slices, each of which \emph{does} make classical sense. All Lorentzian slices that can emerge in JT gravity are slices of the TFD, but at arbitrary times. This is~\eqref{eq:efftime1}. They find the wavefunction by decomposing the full wavefunction into basis functions $e^{-\Delta \ell(T_\text{eff})}$ using~\eqref{eq:efftime1}, as we are doing here.
\item
As already emphasized, we are not claiming that our discussion is directly applicable to describing the experiences of infalling observers. It is generally not well understood how to describe bulk observers in gravitational systems.%
\footnote{For some interesting recent progress, see~\cite{Chandrasekaran:2022cip,Leutheusser:2021qhd}.}
In particular, we do not really know that an infalling observer would experience the effective geometries on the right-hand sides of~\eqref{eq:AT-intro} and \eqref{eq:G-conversion}. Even though this seems plausible, it is an important open problem.
\item The reason to consider contributions to the integrals from $T_\text{eff}\neq 2T$ is that the effect of wormholes introduces the sine kernel in~\eqref{eq:F-final}, which has contributions with very large times like $\sim e^{\pm \i \omega T_H}$. Such Fourier components generate contributions to $\mathcal{F}(T_\text{eff}\rvert 2T)$ when $T_\text{eff}-2T=\pm T_H$. For early times ($T,T_\text{eff}\ll T_H$), the Fourier transform probes the coarse features of $\rho(E_1,E_2)$, and the delta function and sine kernel in~\eqref{eq:F-final} cancel each other out. This is the reason why wormholes are negligible for early time correlators, as they should because our everyday experiences clearly do not involve wormholes. But for very late times $T\sim e^{\S}$, the wormhole corrections \emph{can} compete, and dominate. For instance for two-point functions~\eqref{eq:2point-JT}, the $\rho(E)^2$ contribution will decay to zero in $T\to\infty$ for any $\Delta$ (as there are no infinitely sharp features in the integrand as function of $\omega$). The delta and sine functions in~\eqref{eq:sinekernel} give rise to a non-decaying ramp-plateau contribution~\cite{Cotler:2016fpe,Saad:2018bqo,Saad:2019pqd,Blommaert:2019hjr,Stanford:2022fdt,Blommaert:2022lbh} (which indeed vanishes for $T\ll e^{\S})$.
\end{enumerate}
%

\subsection{Alternative derivation}
\label{sect:quick}

Here we provide a quick derivation for $\mathcal{F}(T_\text{eff}\rvert 2T)$~\eqref{eq:F-final}. The effective-action analysis we just performed will be useful later in section~\ref{sect:3.3}, while the quick derivation here is more similar to the one in section~\ref{sect:3.2}.

Starting with the exact, non-perturbative two-point function~\eqref{eq:2point-JT}, one can write suggestively
\begin{align}
  G_{\Delta\,\text{nonpert}}(2T) = \int^\infty_{-\infty}d \omega\, e^{2\i \omega T}\frac{\rho(E_1,E_2)}{\rho(E_1)\rho(E_2)} G_{\Delta\,\text{class}}(E_1,E_2)\,.
\end{align}
Then we may simply resort to the convolution theorem
\begin{equation}
  \int_{-\infty}^{+\infty}\d \omega\,e^{2\i \omega T}f(\omega)\,h(\omega) = \frac{1}{2\pi} \int_{-\infty}^{+\infty}\d T_{\text{eff}}\, \widetilde{f}(2T-T_\text{eff})\,\widetilde{h}(T_\text{eff})\,,
\end{equation}
where the tilded functions are individual Fourier transforms
\begin{align}
  \widetilde{f}(2T-T_\text{eff}) = \int_{-\infty}^{+\infty}\d \omega\,e^{\i \omega (2T-T_{\text{eff}})}\,f(\omega)\,,\quad\widetilde{h}(2T_\text{eff}) = \int_{-\infty}^{+\infty}\d \omega\,e^{2\i \omega T_{\text{eff}}}\,h(\omega)\,.
\end{align}
Applying this to our case with $f(\omega)=\frac{\rho(E_1,E_2)}{\rho(E_1)\rho(E_2)}$ and $h(\omega)=G_{\Delta\,\text{class}}(E_1,E_2)$ reproduces \eqref{eq:defdistribution} with
\begin{equation}
  \mathcal{F}(T_\text{eff}\rvert 2T)=\frac{1}{2\pi}\int_{-\infty}^{+\infty}\d\omega\,e^{\i \omega (T_\text{eff}-2T)}\bigg(1+\frac{\delta(\omega)}{\rho(E)}-\frac{\sin(\pi\rho(E)\omega)^2}{\pi\omega^2\rho(E)^2}\bigg)\,.\label{eq:243}
\end{equation}
%

\subsection{Gray holes}
\label{sect:gray}

Given the simplified semiclassical expression~\eqref{eq:F-final} for $\mathcal{F}(T_\text{eff}\rvert 2T)$ in the tau-scaling limit, it is simple to compute the all-genus probability of finding an expanding slice $T_\text{eff}>0$ or a contracting slice $T_\text{eff}<0$, as function of boundary time $2T$
\begin{equation}
  P_\text{exp}(2T) = \int_0^{+\infty}\d T_\text{eff}\, \mathcal{F}(T_\text{eff}\rvert 2T)\,,
\quad
  P_\text{cont}(2T) = \int_{-\infty}^{0}\d T_\text{eff}\, \mathcal{F}(T_\text{eff}\rvert 2T)\,.
\label{eq:Pexp-Pcont}
\end{equation}
As an intermediate step, doing the Fourier transform in~\eqref{eq:F-final} leads to a shifted version of the ramp-and plateau structure~\cite{Cotler:2016fpe}
\begin{equation}
  \boxed{\mathcal{F}(T_\text{eff}\rvert 2T) = \delta(T_\text{eff}-2T) + \frac{1}{T_H^2}\text{min}(\abs{T_\text{eff}-2T},T_H)}\,,
\quad
  T_H = 2\pi \rho(E)\,,
\label{eq:F-Fourier}
\end{equation}
which is constant after the Heisenberg~\cite{Haake:1315494} or plateau~\cite{Cotler:2016fpe} time. This results for $2T<T_H$ in the profile
\begin{equation}
  P_\text{exp}(2T) = 1-\frac{2T}{T_H}+\frac{1}{2}\frac{(2T)^2}{T_H^2}+\text{constant}\,,
\quad
  P_\text{cont}(2T) = \frac{2T}{T_H}-\frac{1}{2}\frac{(2T)^2}{T_H^2}+\text{constant}\,,
\label{eq:PP-1}
\end{equation}
and for $2T>T_H$
\begin{equation}
  P_\text{exp}(2T) = \frac{1}{2}+\text{constant}\,,
\quad
  P_\text{cont}(2T) = \frac{1}{2}+\text{constant}\,,
\label{eq:PP-2}
\end{equation}
where the infinite constant (on which we comment soon) is
\begin{equation}
  \text{constant} = \frac{1}{T_H}\int_0^\infty \d T_\text{eff} -\frac{1}{2}\,.
\label{eq:inf-const}
\end{equation}

One noteworthy comment about this result~\eqref{eq:PP-1} is that it agrees at genus one (up to a minus sign) with the results of Stanford and Yang~\cite{Stanford:2022fdt}, who found the quadratic piece $(2T)^2/ 2 T_H^2$. The linear term is geometrically more mysterious, very much like certain terms in the Taylor series in $2T$ of the interior length computed in~\cite{Iliesiu:2021ari}. Even though, as explained in section~\ref{sect:2.1}, our setup is mathematically different from~\cite{Stanford:2022fdt} (due to our treatment of the mapping class group), it is comforting to find this agreement.

Regarding the infinite constant, we believe that the physically correct procedure is to subtract it. More precisely, the amplitude must be \emph{renormalized}, by subtracting the values at $T=0$. Indeed, the definition of our setup is that we start with the TFD at $T=0^+$. Then we ask what happens to this geometry when one time evolves. To interpret $P_\text{exp,\,cont}$ as probabilities, this means we should impose
\begin{align}
  P_\text{exp}(2T=0^+)=1\,, \quad\text{and}\quad P_\text{cont}(2T=0^+)=0\,.
\label{eq:init-cond}
\end{align}
Subtracting this constant boils down to subtracting the ramp-plateau at $2T=0$ in $\mathcal{F}(T_\text{eff}\rvert 2T)$~\eqref{eq:F-Fourier}.%
\footnote{If we use different infrared cutoffs $\Lambda_1$ and $-\Lambda_2$ for the upper and lower edges of the integrals in~\eqref{eq:Pexp-Pcont}, where $\Lambda_{1,2}$ are (much) larger than any other scales in the problem, then the constants for $P_\text{exp}$ and $P_\text{cont}$ in~\eqref{eq:PP-1} and \eqref{eq:PP-2} (which are now regularized) would be different. Even in this case, the condition~\eqref{eq:init-cond} determines the necessary subtractions, so the final result \eqref{2.45bb} is not affected.}
An identical renormalization was carried out when computing the interior length in~\cite{Iliesiu:2021ari}. Nevertheless, this remains a subtle point, one which we will come back to in the discussing section \ref{sect:disclor}.

After this renormalization, one finds that our probabilities are correctly normalized
\begin{equation}
  P_\text{exp}(2T)+P_\text{cont}(2T) = 1
\end{equation}
for all $T$. The resulting piecewise behavior as function of $2T$ makes physical sense
\begin{equation}
    \begin{tikzpicture}[baseline={([yshift=-.5ex]current bounding box.center)}, scale=0.7]
    \pgftext{\includegraphics[scale=1]{kyoto8.pdf}} at (0,0);
    \draw (0,-0.1) node {$1/2$};
    \draw (4.1,-0.7) node {$P_\text{cont}(2T)$};
    \draw (-1.4,-0.7) node {$P_\text{exp}(2T)$};
    \draw (0,1.8) node {$1$};
    \draw (-2.6,-2.8) node {$T_H$};
    \draw (2.9,-2.8) node {$T_H$};
  \end{tikzpicture}\label{2.45bb}
\end{equation}
In particular, expanding and contracting branches plateau at equal odds for post-Heisenberg times
\begin{equation}
  \boxed{P_\text{exp}(2T)=P_\text{cont}(2T)=\frac{1}{2}}\,,\quad 2T > T_H\,.
\end{equation}
This realizes the gray hole scenario anticipated in~\cite{Stanford:2022fdt,Susskind:2015toa}.
For $2T\ll T_H$ wormholes should be irrelevant for all practical purposes, and indeed the geometry is purely expanding, with the transition amplitude only having support on the disk contribution (which gives $T_\text{eff}=2T$ \eqref{cllimit})
\begin{equation}
  P_\text{exp}(2T)=1\,,\quad P_\text{cont}(2T)=0\,,\quad 2T\ll T_H\,.
\end{equation}
We emphasize that in~\eqref{eq:PP-1} the corrections due to wormholes occur at \emph{leading} order. Even though the ramp-plateau in $\mathcal{F}(T_\text{eff}|2T)$ is suppressed by $1/T_H\sim e^{-\S}$ in~\eqref{eq:F-Fourier}, we integrate $T_\text{eff}$ over a range $\sim T_H$ (the length of the ramp) when we compute the probabilities, creating competition at leading order!

As discussed in the introduction, in pure JT gravity, neither expanding nor contracting branch of the pure TFD is dangerous. In the remainder of this work we consider more generic states with early perturbations, which \emph{could} be dangerous.%
\footnote{An even more honest approach would be to treat matter as dynamical quantum fields, thereby including virtual processes such as the vacuum fluctuations. See section~\ref{subsec:improv} for more discussion on this improvement.}

\section{Simple perturbed thermofield double}
\label{sec:one}

In this section, we study the simplest state that may have a dangerous interior, the TFD with a thermal perturbation at $t=-T_w$ on the left boundary. We choose to focus again on exponentially large times $T_w \sim T_H$. The naive dual bulk slice at $T \sim T_H$ (red) is
\begin{equation}
    \begin{tikzpicture}[baseline={([yshift=-.5ex]current bounding box.center)}, scale=0.7]
        \pgftext{\includegraphics[scale=1]{fire5.pdf}} at (0,0);
        \draw (-2.2,1.2) node {$\color{red}T$};
        \draw (-2.2,-0.5) node {$\color{blue}T_w$};
    \end{tikzpicture}
\label{eq:3.1bb}
\end{equation}
The early perturbation creates a shockwave~\cite{Shenker:2013pqa}, which depending on the values of $T$ and $T_\omega$ can be highly disruptive in the slice at time $T$. In particular, in JT gravity the geometry of the slice is characterized by its (renormalized) length $\ell$~\eqref{eq:ell}. For exponentially late times, to leading order in $e^{\S}$, the length of an unperturbed slice is 
\begin{equation}
  \ell_\text{bare}=2 E^{1/2}\abs{2 T}\,.
\label{eq:ellbare}
\end{equation}
In the presence of a shockwave, this may get modified. We \emph{define} a strong shockwave to be one which affects this leading order behavior. We \emph{define} slices with a strong shocks to be dangerous.

At disk level (i.e. ignoring wormholes), the presence of the shock modifies~\eqref{eq:ellbare} to~\cite{Stanford:2014jda}%
\footnote{We will show how this is reproduced from a detailed analysis of the exact JT disk amplitude in section~\ref{sect:3.3}.}
\begin{equation}
  \ell=2E^{1/2}(\abs{T_1}+\abs{T_2})\,,\quad T_1=T-T_w\,,\quad T_2=T+T_w\,.
\label{eq:l-disk-shock}
\end{equation}
Comparing with~\eqref{eq:ellbare}, this leads to the naive conclusion~\cite{Stanford:2014jda} that the slice obeys%
\footnote{The separation between $|T_w|$ and $|T|$ is assumed exponentially large in $\S$.}
\begin{align}
  \abs{T_{w}}&>\abs{T}\quad \text{dangerous}\,,
\nonumber\\
  \abs{T_{w}}&<\abs{T}\quad \text{safe}\,.
\label{eq:regionsintro}
\end{align}
The contribution from wormholes is expected to change this naive conclusion.
In this section, we want to answer the following question:\ including the effects of wormholes, what is the probability that the dual bulk slice is dangerous (meaning it contains a strong shockwave)?

\subsection{Logical overview}
\label{sect:log}

In similar spirit to what we discussed in section~\ref{sect:2.1}, we start our analysis with the boundary four-point function in JT gravity, computed by summing over wormholes as~\cite{Blommaert:2020seb}
\begin{align}
  &G_{\Delta\,\Delta_w\,\text{nonpert}}(T_{1},T_{2})=   \begin{tikzpicture}[baseline={([yshift=-.5ex]current bounding box.center)}, scale=0.7]
        \pgftext{\includegraphics[scale=1]{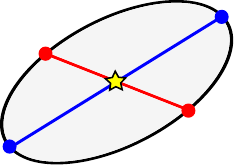}} at (0,0);
        \draw (-2,0.7) node {$\color{red}\mathcal{O}_\Delta$};
        \draw (2,-0.7) node {$\color{red}\mathcal{O}_\Delta$};
        \draw (2.6,0.6) node {$\b_1$};
        \draw (-2.6,-0.6) node {$\b_3$};
        \draw (0,1.7) node {$\b_2$};
        \draw (0,-1.7) node {$\b_4$};
        \draw (-2.6,-1.4) node {$\color{blue}\mathcal{O}_{\Delta_w}$};
        \draw (2.6,1.4) node {$\color{blue}\mathcal{O}_{\Delta_w}$};
    \end{tikzpicture}\quad +\quad \begin{tikzpicture}[baseline={([yshift=-.5ex]current bounding box.center)}, scale=0.7]
        \pgftext{\includegraphics[scale=1]{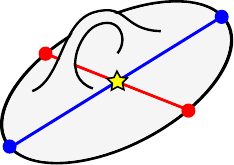}} at (0,0);
        \draw (-2,0.7) node {$\color{red}\mathcal{O}_\Delta$};
        \draw (2,-0.7) node {$\color{red}\mathcal{O}_\Delta$};
        \draw (2.6,0.6) node {$\b_1$};
        \draw (-2.6,-0.6) node {$\b_3$};
        \draw (0,1.7) node {$\b_2$};
        \draw (0,-1.7) node {$\b_4$};
        \draw (-2.6,-1.4) node {$\color{blue}\mathcal{O}_{\Delta_w}$};
        \draw (2.6,1.4) node {$\color{blue}\mathcal{O}_{\Delta_w}$};
    \end{tikzpicture} \nonumber\\
    &\qquad+\quad \begin{tikzpicture}[baseline={([yshift=-.5ex]current bounding box.center)}, scale=0.7]
        \pgftext{\includegraphics[scale=1]{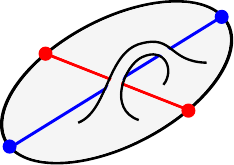}} at (0,0);
        \draw (-2,0.7) node {$\color{red}\mathcal{O}_\Delta$};
        \draw (2,-0.7) node {$\color{red}\mathcal{O}_\Delta$};
        \draw (2.6,0.6) node {$\b_1$};
        \draw (-2.6,-0.6) node {$\b_3$};
        \draw (0,1.7) node {$\b_2$};
        \draw (0,-1.7) node {$\b_4$};
        \draw (-2.6,-1.4) node {$\color{blue}\mathcal{O}_{\Delta_w}$};
        \draw (2.6,1.4) node {$\color{blue}\mathcal{O}_{\Delta_w}$};
    \end{tikzpicture}\quad +\quad \begin{tikzpicture}[baseline={([yshift=-.5ex]current bounding box.center)}, scale=0.7]
        \pgftext{\includegraphics[scale=1]{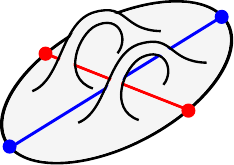}} at (0,0);
        \draw (-2,0.7) node {$\color{red}\mathcal{O}_\Delta$};
        \draw (2,-0.7) node {$\color{red}\mathcal{O}_\Delta$};
        \draw (2.6,0.6) node {$\b_1$};
        \draw (-2.6,-0.6) node {$\b_3$};
        \draw (0,1.7) node {$\b_2$};
        \draw (0,-1.7) node {$\b_4$};
        \draw (-2.6,-1.4) node {$\color{blue}\mathcal{O}_{\Delta_w}$};
        \draw (2.6,1.4) node {$\color{blue}\mathcal{O}_{\Delta_w}$};
    \end{tikzpicture} \quad + \quad \dots
\label{eq:eucl4pt}
\end{align}
with analytic continuation of the boundary conditions to%
\footnote{The genuinely Lorentzian setup~\eqref{eq:3.1bb} has $\alpha=\pi/2$, such that ${\rm Re}\,\beta_1 = {\rm Re}\,\beta_4 = \beta/2$ and ${\rm Re}\,\beta_2 = {\rm Re}\,\beta_3 = 0$. We allow Euclidean separation between all operators, to avoid infinite energies. The precise value of $\alpha$ will not affect our conclusions.
\label{ft:alpha}}
\begin{equation}
  \beta_1=\frac{\beta}{4}+\frac{\beta}{2\pi}\alpha+\i T_1\,,\quad \beta_2=\frac{\beta}{4}-\frac{\beta}{2\pi}\alpha+\i T_2\,,\quad \beta_3=\frac{\beta}{4}-\frac{\beta}{2\pi}\alpha-\i T_2\,,\quad \beta_4=\frac{\beta}{4}+\frac{\beta}{2\pi}\alpha-\i T_1\,.
\label{eq:beta_1-4}
\end{equation}

In \textbf{section~\ref{sect:3.2}}, we show that the out-of-time-order correlator (OTOC) on this contour~\eqref{eq:beta_1-4} exactly decomposes, in the tau-scaling limit, as
\begin{equation}
  G_{\Delta\,\Delta_w\,\text{nonpert}}(T_{1},T_{2}) = \int_{-\infty}^{+\infty}\!\d T_{1\,\text{eff}}\,\mathcal{F}(T_{1\,\text{eff}}\rvert T_1)\int_{-\infty}^{+\infty}\!\d T_{2\,\text{eff}}\,\mathcal{F}(T_{2\,\text{eff}}\rvert T_2)\,G_{\Delta\,\Delta_w\,\text{disk}}(T_{1\,\text{eff}},T_{2\,\text{eff}})\,.
\label{eq:G_4pt-decomp}
\end{equation}
One key step is to show that the dominant wormholes are empty wormholes that bridge over the (red) observer; precisely the ones pictured in~\eqref{eq:eucl4pt}. Other possibilities are wormholes which bridge over the (blue) perturbation; and non-empty wormholes which make the perturbation bypasses the interior slice:
\begin{equation}
    G_{\Delta\,\Delta_w\,\text{nonpert}}(T_{1},T_{2}) \supset \begin{tikzpicture}[baseline={([yshift=-.5ex]current bounding box.center)}, scale=0.7]
        \pgftext{\includegraphics[scale=1]{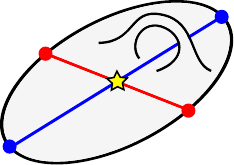}} at (0,0);
        \draw (-2,0.7) node {$\color{red}\mathcal{O}_\Delta$};
        \draw (2,-0.7) node {$\color{red}\mathcal{O}_\Delta$};
        \draw (2.6,0.6) node {$\b_1$};
        \draw (-2.6,-0.6) node {$\b_3$};
        \draw (0,1.7) node {$\b_2$};
        \draw (0,-1.7) node {$\b_4$};
        \draw (-2.6,-1.4) node {$\color{blue}\mathcal{O}_{\Delta_w}$};
        \draw (2.6,1.4) node {$\color{blue}\mathcal{O}_{\Delta_w}$};
    \end{tikzpicture}\quad + \quad \begin{tikzpicture}[baseline={([yshift=-.5ex]current bounding box.center)}, scale=0.7]
        \pgftext{\includegraphics[scale=1]{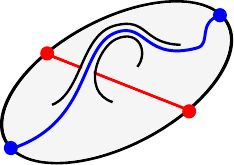}} at (0,0);
        \draw (-2,0.7) node {$\color{red}\mathcal{O}_\Delta$};
        \draw (2,-0.7) node {$\color{red}\mathcal{O}_\Delta$};
        \draw (2.6,0.6) node {$\b_1$};
        \draw (-2.6,-0.6) node {$\b_3$};
        \draw (0,1.7) node {$\b_2$};
        \draw (0,-1.7) node {$\b_4$};
        \draw (-2.6,-1.4) node {$\color{blue}\mathcal{O}_{\Delta_w}$};
        \draw (2.6,1.4) node {$\color{blue}\mathcal{O}_{\Delta_w}$};
    \end{tikzpicture}\quad + \quad \dots
\label{eq:G_4pt-wormhole}
\end{equation}
We find that these are \emph{negligible} in the tau-scaling limit~\eqref{eq:tauscaling}. To be clear, every wormhole is exponentially suppressed in entropy by a factor of $e^{-2 \S}$, but some are amplified at exponentially late times due to the integrals over moduli $T_{i\,\text{eff}}$. The second term in~\eqref{eq:G_4pt-wormhole} is the ``safest'' geometry (the perturbation bypasses the observer). This term, however, does not have such moduli, and hence no enhancement in time. The kernels appearing in~\eqref{eq:G_4pt-decomp} notably have precisely the same universal form as~\eqref{eq:F-Fourier}
\begin{equation}
  \mathcal{F}(T_{i\,\text{eff}}\rvert T_i) = \delta(T_{i\,\text{eff}}-T_i) + \frac{1}{T_H^2}\text{min}(\abs{T_{i\,\text{eff}}-T_i},T_H)\,,\quad T_H=2\pi \rho(E)\,.
\end{equation}

An important point is that, unlike the previous case where $G_{\Delta\,\text{class}} = e^{-\Delta \ell}$, the OTOC, even at the disk level, does \emph{not} always factorize
\begin{equation}
    G_{\Delta\,\Delta_w\,\text{disk}}(T_{1\,\text{eff}},T_{2\,\text{eff}})=   \begin{tikzpicture}[baseline={([yshift=-.5ex]current bounding box.center)}, scale=0.7]
        \pgftext{\includegraphics[scale=1]{fire15.pdf}} at (0,0);
        \draw (-2,0.7) node {$\color{red}\mathcal{O}_\Delta$};
        \draw (2,-0.7) node {$\color{red}\mathcal{O}_\Delta$};
        \draw (2.6,0.6) node {$\b_1$};
        \draw (-2.6,-0.6) node {$\b_3$};
        \draw (0,1.7) node {$\b_2$};
        \draw (0,-1.7) node {$\b_4$};
        \draw (-2.6,-1.4) node {$\color{blue}\mathcal{O}_{\Delta_w}$};
        \draw (2.6,1.4) node {$\color{blue}\mathcal{O}_{\Delta_w}$};
    \end{tikzpicture} \neq e^{-\Delta\ell}e^{-\Delta_w\ell_{w}},
\end{equation}
where $\ell_w$ is the renormalized geodesic length between the two $\mathcal{O}_{\Delta_w}$ insertions. The correct answer is given by the Schwarzian (or disk) OTOC four-point function (for fixed energy), computed for instance in~\cite{Mertens:2017mtv,Blommaert:2018oro} (and semiclassically in~\cite{Maldacena:2016upp}). Even though this OTOC has been well studied in the literature, we conduct a more careful semiclassical (``scramblon'' \cite{Stanford:2021bhl}) analyses in \textbf{section~\ref{sect:3.3}} to decompose it as
\begin{equation}
  G_{\Delta\,\Delta_w\,\text{disk}}(T_{1\,\text{eff}},T_{2\,\text{eff}})=\int_{-\infty}^{+\infty}\!\d\ell\,\mathcal{A}_{\Delta_w\,\text{disk}}(T_{1\,\text{eff}},T_{2\,\text{eff}}, \ell)\,e^{-\Delta \ell}\,,
\label{eq:A_Teff_l}
\end{equation}
where $\ell$ is the length of the interior slice. As in \eqref{eq:Al_def} we identify the kernel in \eqref{eq:A_Teff_l} as the amplitude that prepares the slice with length $\ell$ (and one perturbation). This results in the relation%
\footnote{We normalize this amplitude as follow. Imagine that there are no insertion $O_\Delta$ at the time slice $T$. Then the amplitude is equal to the two-point function of the perturbation $ \langle O_{\Delta_w}(\beta_2+\beta_3) O_{\Delta_w}(0)\rangle$. We divide the amplitude by this two-point function so that it is normalized to one. In~\eqref{eq:4pt-lengths}, we will confirm that this normalization is indeed time-independent.\label{fn:20}}
\begin{equation}
  \mathcal{A}_{\Delta_w\,\text{nonpert}}(T_1,T_2,\ell)=\int_{-\infty}^{+\infty}\!\d T_{1\,\text{eff}}\,\mathcal{F}(T_{1\,\text{eff}}\rvert T_1)\int_{-\infty}^{+\infty}\!\d T_{2\,\text{eff}}\,\mathcal{F}(T_{2\,\text{eff}}\rvert T_2)\,\mathcal{A}_{\Delta_w\,\text{disk}}(T_{1\,\text{eff}},T_{2\,\text{eff}}, \ell)\,.
\label{eq:ampbb}
\end{equation}
This leads to equation~\eqref{eq:G-conversion} in the introduction, with $T_1=T-T_w$ and $T_2=T+T_w$
\begin{align}
    &\begin{tikzpicture}[baseline={([yshift=-.5ex]current bounding box.center)}, scale=0.7]
        \pgftext{\includegraphics[scale=1]{fire5.pdf}} at (0,0);
        \draw (-2.2,1.2) node {$\color{red}T$};
        \draw (-2.2,-0.5) node {$\color{blue}T_w$};
    \end{tikzpicture}\quad+\text{ wormholes }\nonumber\\&\qquad\qquad\qquad=\int_{-\infty}^{+\infty}\d T_{1\,\text{eff}}\,\mathcal{F}(T_{1\,\text{eff}}\rvert T_1)\int_{-\infty}^{+\infty}\d T_{2\,\text{eff}}\,\mathcal{F}(T_{2\,\text{eff}}\rvert T_2)\,
    \begin{tikzpicture}[baseline={([yshift=-.5ex]current bounding box.center)}, scale=0.7]
        \pgftext{\includegraphics[scale=1]{fire6.pdf}} at (0,0);
        \draw (-2.3,0.4) node {$\color{red}T_\text{eff}$};
        \draw (-2.5,-1.2) node {$\color{blue}T_{w\,\text{eff}}$};
    \end{tikzpicture}\,\,\,.
\label{eq:4pt-exp}
\end{align}

The final step is to figure out what ``safe'' and ``dangerous'' means in the effective, classical geometry.
In \textbf{section~\ref{sect:3.3}}, starting from the exact disk OTOC in JT gravity, we show that the normalized version of $\mathcal{A}_{\Delta_w\,\text{disk}}(T_{1\,\text{eff}},T_{2\,\text{eff}},\ell)$ simply reproduces the classical expectation of~\cite{Stanford:2014jda}, with $T_{1\,\text{eff}}=T_\text{eff}-T_{w\,\text{eff}}$ and $T_{2\,\text{eff}}=T_\text{eff}+T_{w\,\text{eff}}$
\begin{equation}
  \mathcal{A}_{\Delta_w\,\text{disk}}(T_{1\,\text{eff}},T_{2\,\text{eff}},\ell) = \delta\bigl(\ell-2E^{1/2}(\abs{T_{1\,\text{eff}}}+\abs{T_{2\,\text{eff}}})\bigr)\,.
\label{eq:A-delta}
\end{equation}
Following \eqref{eq:regionsintro}, we conclude from~\eqref{eq:A-delta} that the effective classical bulk slice being probed is dangerous respectively safe if
\begin{align}
  \boxed{\begin{aligned}\abs{T_{w\,\text{eff}}}&>\abs{T_\text{eff}}\quad \text{dangerous}\\\abs{T_{w\,\text{eff}}}&<\abs{T_\text{eff}}\quad \text{safe}\end{aligned}}\,\,.
\label{eq:regions-disk-eff}
\end{align}

Putting everything together, in \textbf{section~\ref{survival}} we compute the probability that the dual bulk slice is safe/dangerous, by integrating~\eqref{eq:ampbb} over the regions~\eqref{eq:regions-disk-eff}. The $\ell$ integral is trivial, due to the delta function~\eqref{eq:A-delta}, resulting in
\begin{equation}
  P_\text{safe}(T_1,T_2) = \int_{\text{safe}}\!\d T_{1\,\text{eff}}\,\d T_{2\,\text{eff}}\,\mathcal{F}(T_{1\,\text{eff}}\rvert T_1)\mathcal{F}(T_{2\,\text{eff}}\rvert T_2)\,.
\label{eq:P_safe-T1-T2}
\end{equation}
As an aside, we note that there might be other sensible criteria besides~\eqref{eq:regionsintro} to distinguish whether or not the slice is dangerous. Our techniques allow for extending any replacement of the classical criterion~\eqref{eq:regionsintro} to the corresponding non-perturbative criterion (and resulting safe/danger probabilities) in a straightforward manner.

In \textbf{section~\ref{sect:folds}}, we rephrase \eqref{eq:regions-disk-eff} using effective time-folds, as advertised in~\eqref{1.4}. This results in
\begin{align}
  \boxed{\begin{aligned}\text{effective out-of-time-ordered}\quad &\text{ dangerous}\\\text{effective time-ordered}\quad & \text{ safe}\end{aligned}}\,\,,
\label{eq:regions-eff-tf}
\end{align}
which, hopefully, is more intuitive. We emphasize that it is the \emph{effective} time-folds that are in one-to-one correspondence with the \emph{classical spacetime}, not the boundary time-folds (which serve as the boundary condition for the exact calculation).

Sections~\ref{sect:3.2} and \ref{sect:3.3} are technical. Readers interested only in physical results can skip to section~\ref{survival}.

\subsection{Decomposing exact amplitude using effective times}
\label{sect:3.2}

Except for contributions such as the last diagram in~\eqref{eq:G_4pt-wormhole} (which we address in section~\ref{survival}), the exact four-point function in JT gravity~\eqref{eq:eucl4pt} decomposes as~\cite{Blommaert:2020seb,Yang:2018gdb}
\begin{align}
  &G_{\Delta\,\Delta_w\,\text{nonpert}}(T_{1},T_{2})
\label{eq:4ptex}\\
  &=\prod_{i=1}^4\int_0^\infty \frac{\d z_i}{z_i}\,\int_0^\infty\! \d E_i\,\psi_{E_i}(z_i)\,e^{-\beta_i E_i}\,\rho(E_1,E_2,E_3,E_4) \int_0^\infty\! \frac{\d z}{z}\,I_3(z_1,z_2,z)I_3(z_3,z_4,z)\,e^{-\Delta_w \ell_w} e^{-\Delta \ell}\,.
\nonumber
\end{align}
Here, we introduced the notation
\begin{equation}
  z_i=e^{-\ell_i/2}\,,\quad z=e^{-\ell/2}\,,
\end{equation}
where $\ell_i$ is the renormalized geodesic length between the $i$-th and $(i+1)$-the insertions on the boundary. The four-level spectral density $\rho(E_1,E_2,E_3,E_4)$ arises from the sum over wormholes and should be computed in the tau-scaling limit as the generalization of~\eqref{eq:sinekernel}. Furthermore,
\begin{equation}
  I_3(z_1,z_2,z_3)=\int\! \d E\,\rho_0(E)\,\prod_{i=1}^3\psi_E(z_i)=\exp\bigg(-\frac{1}{2}\frac{z_1 z_2}{z_3}-\frac{1}{2}\frac{z_2 z_3}{z_1}-\frac{1}{2}\frac{z_3 z_1}{z_2}\bigg)\,.
\label{eq:I_3}
\end{equation}
Finally, the length $\ell_w$ of the geodesic of the perturbation follows from some hyperbolic geometry
\begin{equation}
  e^{\ell_w/2}=\frac{1}{z_w}=\frac{z}{z_1 z_3}+\frac{z}{z_2 z_4}\,.
\end{equation}
The derivation is somewhat elaborate, but well documented. We here only summarize the main steps leading to~\eqref{eq:4ptex} in words, and refer the interested reader to the relevant literature for details.
\begin{enumerate}
\item
By using the correspondence between JT gravity and 2d topological Yang-Mills theory (also called BF theory), and thinking carefully about the mapping class group, one derives a version of~\eqref{eq:4ptex} that, instead of all the $z_i$ and $z$ dependence, has a kernel which (asides from the four-level spectral density) involves a 6j~symbol of SL$(2,\mathbb{R})$ and some gamma functions~\cite{Blommaert:2020seb}.
\item
It remains to show that the integrals over $z_i$ and $z$ reproduce the said 6j~symbol and gamma functions. This can be done by focusing (as a technical tool) on the expression for the JT disk four-point function, which is identical to~\eqref{eq:4ptex} except that the four-level spectral density is replaced by the product of 4~disk spectral densities $\rho_0(E_1)\dots \rho_0(E_4)$.

The four-point function with a 6j~symbol and gamma functions follows from various perspectives on the quantization of JT gravity~\cite{Mertens:2017mtv,Blommaert:2018oro,Blommaert:2018iqz,Iliesiu:2019xuh}. The version with the $z_i$ and $z$ integrals follows from a direct quantization of the dual Schwarzian quantum mechanics~\cite{Yang:2018gdb,Kolchmeyer:2023gwa}. In particular, the propagator is the wavefunction $\psi_{E_i}(z_i)$ multiplied by a ``phase,'' and those phases combine into~\eqref{eq:I_3}. The factors $e^{-\Delta_w \ell_w}\,e^{-\Delta \ell}$ are just the OTOC Schwarzian bilocals written out in this propagator language~\cite{Yang:2018gdb}. These two quantizations of the same theory prove the relation which we wanted.
\end{enumerate}

Let us now label the four energies as
\begin{align}
  E_1&=E+\frac{\bar{\omega}}{4}+\frac{\omega_1}{2}\,,\quad E_2=E-\frac{\bar{\omega}}{4}+\frac{\omega_2}{2}\,,\quad E_3=E-\frac{\bar{\omega}}{4}-\frac{\omega_2}{2}\,,\quad E_4=E+\frac{\bar{\omega}}{4}-\frac{\omega_1}{2}\,.
\label{eq:energies}
\end{align}
Here, $\omega_1$ (resp. $\omega_2$) is the energy difference between $E_1$ and $E_4$ (resp. $E_2$ and $E_3$), which will be exponentially small. On the other hand, $\bar{\omega}$, will be $O(1)$ (details follow). Introducing an $\alpha$ parameter as in~\eqref{eq:beta_1-4}, with
\begin{equation}
  a=\frac{\b}{2\pi}\a\,,
\end{equation}
the Boltzmann weights in~\eqref{eq:4ptex} combine into $e^{-\beta E-a \bar{\omega}}e^{-\i \omega_1 T_1 - \i \omega_2 T_2}$. We again work at fixed energy $E$
\begin{equation}
  \frac{\b}{2\pi}=\frac{1}{2E^{1/2}}\,.
\end{equation}
Below, this is always the meaning of $\b$. Then we can define an angular variable at fixed energy, which we will again call $\alpha$, and its Legendre dual frequency $\omega$ as follows
\begin{equation}
  \a=2 E^{1/2} a\,,\quad \omega = \frac{\bar{\omega}}{2 E^{1/2}}\,.
\end{equation}
This leaves us with the rewriting of~\eqref{eq:4ptex}
\begin{align}
  &G_{\Delta\,\Delta_w\,\text{nonpert}}(T_{1},T_{2})\nonumber\\&=\int_{-\infty}^{+\infty}\! \d\omega\,e^{-\alpha \omega}\int_{-\infty}^{+\infty}\!\d\omega_1\,e^{-\i \omega_1 T_1}\int_{-\infty}^{+\infty}\!\d\omega_2\,e^{-\i \omega_2 T_2}\,\frac{\rho(E_1,E_2,E_3,E_4)}{\rho_0(E_1)\dots \rho_0(E_4)}\,G_{\Delta\,\Delta_w\,\text{disk}}(E_1,E_2,E_3,E_4)\,,
\label{eq:G_int-rep}
\end{align}
where
\begin{equation}
  G_{\Delta\,\Delta_w\,\text{disk}}(E_1,E_2,E_3,E_4)=\prod_{i=1}^4\int_0^\infty\! \frac{\d z_i}{z_i}\,\rho_0(E_i)\psi_{E_i}(z_i)\int_0^\infty\! \frac{\d z}{z}\,I_3(z_1,z_2,z)I_3(z_3,z_4,z)\,e^{-\Delta_w \ell_w}\,e^{-\Delta \ell}\,.\label{eq:327}
\end{equation}
We are interested in this triple Fourier transform, \eqref{eq:G_int-rep}, in the \emph{tau scaling limit}
\begin{equation}
  T_1\,,T_2\,,e^{\S}\to \infty \,,\quad T_1 e^{-\S}\,,T_2 e^{-\S}\text{ fixed}.
\label{eq:tau-scaling_4pt}
\end{equation}
As explained around~\eqref{eq:tau-scaling_omega}, this corresponds (as announced before) to considering $\omega_1, \omega_2 \sim e^{-\S}$.

The integral~\eqref{eq:G_int-rep} gets most of its meaningful contributions (because we consider $\alpha \sim 1$) from $\omega \sim 1$. This means that for the overwhelming majority of the integrand we are considering $(E_1,E_4)-(E_2,E_3)\gg e^{-\S}$. In other words, in the limit~\eqref{eq:tau-scaling_4pt} one should consider $E_1-E_4\sim e^{-\S}$ and $E_2-E_3\sim e^{-\S}$, but all other energy differences much \emph{larger} than the inverse level spacing $\sim e^{-\S}$. In this regime, the exact random matrix theory answer factorizes~\cite{mehta2004random}
\begin{equation}
  \frac{\rho(E_1\dots E_4)}{\prod_{i=1}^4 \rho_0(E_i)}=\bigg(1+\frac{\delta(\omega_1)}{\rho(E)}-\frac{\sin(\pi\rho(E)\omega_1)^2}{\pi\rho(E)^2\omega_1^2}\bigg)\bigg(1+\frac{\delta(\omega_2)}{\rho(E)}-\frac{\sin(\pi\rho(E)\omega_2)^2}{\pi\rho(E)^2\omega_2^2}\bigg)\,.
\label{eq:rho-factor}
\end{equation}
This corresponds to the statement that wormholes that ``bridge over'' the perturbation in~\eqref{eq:G_4pt-wormhole} may be ignored in the tau scaling limit.%
\footnote{More generally, the intuition is that wormholes over a line are only important when there is a large and opposite time on each side of the line.}
Note that in this regime we may treat the kernel as independent of $\omega$. Finally, to arrive at~\eqref{eq:G_4pt-decomp} we simply use as in section~\ref{sect:quick} the convolution theorem
\begin{equation}
  \int_{-\infty}^{+\infty}\!\d \omega_1\,e^{\i \omega_1 T_1}f(\omega_1)h(\omega_1)=\frac{1}{2\pi}\int_{-\infty}^{+\infty}\!\d T_{1\,\text{eff}} \int_{-\infty}^{+\infty}\!\d \omega_a\,e^{\i \omega_a (T_1-T_{1\,\text{eff}})}\,f(\omega_a)\int_{-\infty}^{+\infty}\!\d \omega_b\,e^{\i \omega_b T_{1\,\text{eff}}}\,h(\omega_b).
\end{equation}
Using~\eqref{eq:243} for the Fourier transform of~\eqref{eq:rho-factor} and applying the convolution theorem to~\eqref{eq:G_int-rep} gives%
\footnote{We could have used the convolution theorem also for the $\omega$ dependence. However, even in our decomposition of the state~\eqref{eq:ampbb} there is always the factor $e^{-\Delta_w \ell_w}$, which quickly decays to zero for large Lorentzian times. Then we are left with doing a short-Lorentzian time $\omega$ integral of the four-level spectral density. This is exponentially dominated by the contributions without wormholes bridging over the blue line, \eqref{eq:rho-factor}, bringing us back to the situation we present here.}
\begin{equation}
    G_{\Delta\,\Delta_w\,\text{nonpert}}(T_{1},T_{2})=\int_{-\infty}^{+\infty}\d T_{1\,\text{eff}}\,\mathcal{F}(T_{1\,\text{eff}}\rvert T_1)\int_{-\infty}^{+\infty}\d T_{2\,\text{eff}}\,\mathcal{F}(T_{2\,\text{eff}}\rvert T_2)\,G_{\Delta\,\Delta_w\,\text{disk}}(T_{1\,\text{eff}},T_{2\,\text{eff}})\,,
\label{eq:G_T1T2-decomp}
\end{equation}
with
\begin{equation}
  G_{\Delta\,\Delta_w\,\text{disk}}(T_{1},T_{2})=\int_{-\infty}^{+\infty}\! \d\omega\,e^{-\alpha \omega}\int_{-\infty}^{+\infty}\!\d\omega_1\,e^{-\i \omega_1 T_1}\int_{-\infty}^{+\infty}\!\d\omega_2\,e^{-\i \omega_2 T_2}\,G_{\Delta\,\Delta_w\,\text{disk}}(E_1,E_2,E_3,E_4)\,.
\end{equation}
%

\subsection{Semiclassical (scramblon) analysis of disk amplitude}
\label{sect:3.3}

We are interested in the semiclassical interpretation of the OTOC $G_{\Delta\,\Delta_w\,\text{disk}}(T_{1\,\text{eff}},T_{2\,\text{eff}})$ and its decomposition~\eqref{eq:A_Teff_l}. In this subsection, we will discuss exclusively the correlation function on the effective geometry, so we will drop the subscript ``eff'' from here on. We will follow closely~\cite{Stanford:2021bhl}, who adopt a very geometric way of thinking about the disk amplitude via ``scramblons,''%
\footnote{Several other semiclassical descriptions of this OTOC are available~\cite{Maldacena:2016upp,Lam:2018pvp,Stanford:2021bhl}, each with their own value.}
The material of section~\ref{sect:semiwave} prepares for the current discussion.%
\footnote{The results of section~\ref{sect:3.2} can in fact be reproduced (along the lines of section~\ref{sect:semiwave}) by considering off-shell values of $\alpha_i$ that correspond to the effective times by $\alpha_i=2\pi \beta_{i\,\text{eff}}/\beta$. We will not present this alternative derivation.}
We start with repeating the disk four-point function~\eqref{eq:327}
\begin{equation}
  \prod_{i=1}^4\int_0^\infty\! \frac{\d z_i}{z_i}\,\int_0^\infty\! \d E_i\,\psi_{E_i}(z_i)\,e^{-N \beta_i E_i}\,\rho_0(E_i)\int_0^\infty\! \frac{\d z}{z}\, I_3(z_1,z_2,z) I_3(z_3,z_4,z)\, e^{-\Delta_w \ell_w}\, e^{-\Delta \ell}\,.
\label{eq:A_disk}
\end{equation}
One can think of this as associated with a particle propagating near the boundary of AdS$_2$~\cite{Yang:2018gdb}. The integration variables $z_i$ and $z$ parameterize the locations along the particle's trajectory on AdS$_2$ where operators are inserted. Using the 3~degrees of freedom of the SL$(2,\mathbb{R})$ isometry of AdS$_2$, the 8~degrees of freedom parameterizing the locations of the 4~operators on AdS$_2$ reduce to $5$ physical coordinates $(z_i,z)$. Parameterizing AdS$_2$ as
\begin{equation}
  \d s^2 = \d\rho^2 +\sinh^2(\rho+2\log(\Phi_b/\varepsilon))\,\d\theta^2\,,
\label{eq:coord}
\end{equation}
the renormalized geodesics length $\ell_{ij}$ between the $i$-th and the $j$-th boundary locations is%
\footnote{It should be understood that $0<\theta_i-\theta_j<2\pi$ such that the length $\ell_{ij}$ is always positive.}
\begin{equation}
  z_{i j} = e^{-\ell_{ij}/2} = 4\frac{e^{-\frac{\rho_i+\rho_j}{2}}}
    {\sin \frac{\theta_i-\theta_j}{2}}\,.
\end{equation}
A general off-shell contribution to the`` particle propagator''~\eqref{eq:A_disk} looks like~\cite{Yang:2018gdb,Maldacena:2016upp}
\begin{equation}
     \begin{tikzpicture}[baseline={([yshift=-.5ex]current bounding box.center)}, scale=0.7]
        \pgftext{\includegraphics[scale=1]{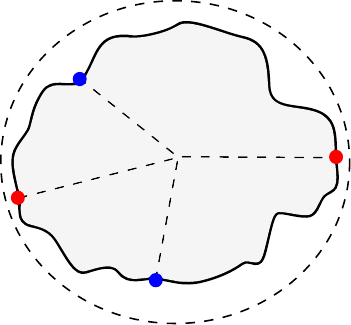}} at (0,0);
         \draw (0.6,1) node {$\theta_1$};
         \draw (0.8,-0.9) node {$\theta_4$};
         \draw (-1.5,0.4) node {$\theta_2$};
         \draw (-1,-1) node {$\theta_3$};
         \draw (3.5,0) node {$\color{red}\rho$};
         \draw (-2.55,2.25) node {$\color{blue}\rho_w$};
    \end{tikzpicture}
\label{eq:disk-pic}
\end{equation}
The sum of the angles spanned by the geodesics add up to $2\pi$ (even off-shell)
\begin{equation}
  \sum_{i=1}^4\theta_i=2\pi\,.
\end{equation}
Combining 3~independent angles with the radial location $\rho$ and $\rho_w$ of the operator insertions $\mathcal{O}_{\Delta}$ and $\mathcal{O}_{\Delta_w}$ indeed gives 5~independent coordinates.%
\footnote{The SL$(2,\mathbb{R})$ isometry was used as follows. Two translations were used to place the two $\mathcal{O}_{\Delta}$ at the same radial coordinate, \emph{ditto} for the $\mathcal{O}_{\Delta_w}$ insertions. The rotation then may be used to put the first insertion at $\theta=0$.}

Following~\cite{Stanford:2021bhl}, we introduced $\Phi_b = N/2$ in~\eqref{eq:coord}, which makes it obvious how to take a classical limit of~\eqref{eq:A_disk}.%
\footnote{Here $N$ plays the role of Newton's constant $G_\text{N}$ (hence large in the semiclassical limit). This is in addition to the small $\varepsilon$ (the IR cutoff of AdS$_2$) limit. Introducing $N$ is not essential. Most JT references use $N=1$, and we do this throughout this work, except in this section. The semiclassical limit is usually obtained by taking $\beta_i$ small and shifting $\rho$ according to~\eqref{eq:coord} to make the action large. Indeed, JT amplitudes only depend on the ratio $\beta_i/2\Phi_b$, which in most early literature was called $\beta_i/C$~\cite{Maldacena:2016upp,Engelsoy:2016xyb,Mertens:2017mtv,Lam:2018pvp}.}
We have also rescaled energies with a factor $N^2$ for transparency. Then labeling energies as in~\eqref{eq:energies}, imposing $\omega,\omega_1,\omega_2\ll E$, and using the integral representations of the Bessel functions as in~\eqref{eq:integral-rep} and \eqref{eq:alpha_i} (this introduces the $\alpha_i$ variables), one finds the ``action'' for the amplitude~\eqref{eq:A_disk}
\begin{align}
  &\prod_{i=1}^4\int \frac{\d z_i}{z_i}\frac{\d z}{z}\,\psi_{E_i}(z_i)\,\rho_0(E_i)\,I_3(z_1,z_2,z)I_3(z_3,z_4,z)
\nonumber\\
  &\overset{\text{class}}{=}\int\! \d\lambda \prod_{i=1}^4\d\a_i\,\d\theta_i\,\d\rho\,\d \rho_w \exp\bigg\lbrace N \bigg( E^{1/2}\sum_{i=1}^4\a_i +\frac{\omega}{4}(  \a_1-\a_2-\a_3+\a_4 )+\frac{\omega_1}{4 E^{1/2}}(\a_1-\a_4)
\nonumber\\
  &\qquad\qquad+\frac{\omega_2}{4 E^{1/2}}(\a_2-\a_3)+4\sum_{i=1}^4e^{-\frac{\rho+\rho_w}{2}}\frac{\cos(\a_i/2)}{\sin(\theta_i/2)}-2\sum_{i=1}^4\frac{e^{-\rho}+e^{-\rho_w}}{\tan(\theta_i/2)}+\lambda \bigg\lbrace\sum_{i=1}^4\theta_i-2\pi\bigg\rbrace\bigg)\bigg\rbrace\,.
\label{eq:3.38ac}
\end{align}
This is to be understood as an equality at the level of the action. We will not attempt to track one-loop factors (including integration measures and contours) in this illustrative calculation. Indeed, the final answer~\eqref{eq:kummer} of this calculation is well known (including the correct one-loop factors)~\cite{Lam:2018pvp}. We merely want to illustrate the geometric interpretation of the underlying calculation.

Because of the large factor $N\gg 1$ up front in~\eqref{eq:3.38ac}, one can do all of the integrals by saddle point. Keeping in mind the Boltzmann weight in~\eqref{eq:G_int-rep}, the $\omega,\omega_1,\omega_2$ integrals localize on
\begin{equation}
  4\a=\a_1-\a_2-\a_3+\a_4\,,\quad 4\i E^{1/2}T_1=\alpha_1-\alpha_4\,,\quad 4\i E^{1/2}T_2=\alpha_2-\alpha_3\,,
\label{eq:integ-local}
\end{equation}
leaving only the sum of $\alpha_i$ unfixed. More generally the equations of motion are
\begin{align}
  0&=E_i^{1/2}-2e^{-\frac{\rho+\rho_w}{2}}\frac{\sin(\a_i/2)}{\sin(\theta_i/2)}\,,
\nonumber\\
  0&=\lambda-2e^{-\frac{\rho+\rho_w}{2}}\frac{\cos(\a_i/2)\cos(\theta_i/2)}{\sin^2(\theta_i/2)}+\frac{e^{-\rho}+e^{-\rho_w}}{\sin^2(\theta_i/2)}\,,
\nonumber\\
  0&=e^{-\frac{\rho}{2}}\sum_{i=1}^4 \frac{\cos(\a_i/2)}{\sin(\theta_i/2)}-e^{-\frac{\rho_w}{2}}\sum_{i=1}^4\frac{1}{\tan(\theta_i/2)}\,,
\nonumber\\
  0&=e^{-\frac{\rho_w}{2}}\sum_{i=1}^4 \frac{\cos(\a_i/2)}{\sin(\theta_i/2)}-e^{-\frac{\rho}{2}}\sum_{i=1}^4\frac{1}{\tan(\theta_i/2)}\,,
\nonumber\\0&=\sum_{i=1}^4\theta_i-2\pi\,.
\end{align}
It is obvious from these equations that the unique classical solution is
\begin{equation}
  \a_i=\theta_i\,,\quad\omega=\omega_1=\omega_2=0\,,\quad e^{-\rho}=e^{-\rho_w}=\frac{E^{1/2}}{2}\,,\quad \lambda=-E^{1/2}\,.
\end{equation}
The first equation fixes the sum of the $\alpha_i$ to $2\pi$, so combined with~\eqref{eq:integ-local}
\begin{equation}
  \a_1=\frac{\pi}{2}+\a+\i \frac{2\pi}{\b} T_1\,,\quad \a_2=\frac{\pi}{2}-\a+\i \frac{2\pi}{\b} T_2\,,\quad \a_3=\frac{\pi}{2}-\a-\i \frac{2\pi}{\b} T_2\,,\quad \a_4=\frac{\pi}{2}+\a-\i \frac{2\pi}{\b} T_1\,,
\label{eq:cl-alpha_i}
\end{equation}
which can be summarized as the fact that $\alpha_i$ are (on-shell) fractions of the boundary length (see~\eqref{eq:alpha_i-disk})
\begin{equation}
  \alpha_i = \frac{2\pi}{\beta}\,\beta_i\,.
\end{equation}
The on-shell action only comes from the first term in the exponent of~\eqref{eq:3.38ac} and equals the entropy in JT gravity $2\pi E^{1/2}$~\eqref{eq:bessels}. Geometrically, this saddle is obvious. The extremum of~\eqref{eq:disk-pic} is simply
\begin{equation}
     \begin{tikzpicture}[baseline={([yshift=-.5ex]current bounding box.center)}, scale=0.7]
        \pgftext{\includegraphics[scale=1]{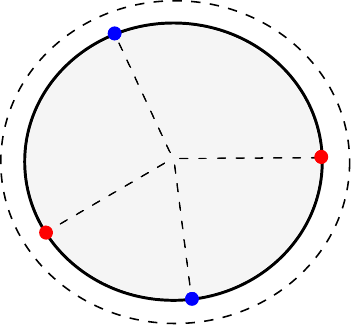}} at (0,0);
        \draw (0.6,1) node {$\a_1$};
         \draw (1.1,-0.8) node {$\a_4$};
         \draw (-1.5,0.4) node {$\a_2$};
         \draw (-0.6,-1) node {$\a_3$};
         \draw (3.5,0) node {$\color{red}\rho$};
         \draw (-2.55,2.25) node {$\color{blue}\rho_w$};
    \end{tikzpicture}
\end{equation}
The classical values $\ell_\text{bare}$ and $\ell_{w\,\text{bare}}$ of the lengths of the geodesics between the two $\mathcal{O}_\Delta$ and the two $\mathcal{O}_{\Delta_w}$ operators, respectively, can be computed from this classical geometry as
\begin{equation}
  e^{-\ell_\text{bare}/2}=\frac{2 E^{1/2}}{\cosh(E^{1/2}(T_1+T_2))}\,,\quad e^{-\ell_{w\,\text{bare}}/2}= \frac{2 E^{1/2}}{\cos(\a)}\,.
\label{eq:4pt-lengths}
\end{equation}
$\ell_\text{bare}$ matches with~\eqref{eq:ell}, up to an additive constant due to slightly different renormalization schemes.%
\footnote{$\ell_{w\,\text{bare}}$ shows that one should regulate the $\mathcal{O}_{\Delta_w}$ insertions with a Euclidean time separation $\alpha\neq \pi/2$ as in footnote~\ref{ft:alpha}; otherwise, $e^{-\Delta_w\ell_w}$ would blow up and backreact heavily on this classical geometry (the integrals in~\eqref{eq:A_disk}). Also notice that $e^{-\Delta_w\ell_{w\,\text{bare}}}$ is \emph{time-independent}, so it makes sense to divide it out as a normalization, as discussed around~\eqref{eq:ampbb}.}

Naively the large $N\gg 1$ factor in~\eqref{eq:3.38ac} guarantees a sharp saddle. However, it can occur that this saddle-point approximation breaks down~\cite{Stanford:2021bhl}. This happens when $e^{-2 E^{1/2}T_i} N=O(1)$, which is at the scrambling time~\cite{Shenker:2013pqa}. In particular, it turns out that two of the dimensions in~\eqref{eq:3.38ac} do not have sharp extrema anymore in this case~\cite{Stanford:2021bhl}, so that one should do the honest integral over those soft directions.%
\footnote{This is familiar from spontaneous symmetry breaking. The soft modes are the pseudo Nambu--Goldstone modes.}
In our setup the soft modes, or ``scramblons''~\cite{Maldacena:2016upp,Stanford:2021bhl,Stanford:2023npy}, capture Dray--'t~Hooft shockwave interactions in the bulk. We will now see how this happens here.

On the classical saddle~\eqref{eq:cl-alpha_i}, both $\alpha_i$ and $\theta_i$ have large imaginary parts for large Lorentzian times. We are interested in configurations close to those saddles. For $T_1$ and $T_2$ to be positive and of order the scrambling time or larger, the action~\eqref{eq:3.38ac} in this regime becomes
\begin{equation}
  \exp\bigg\lbrace N\bigg( \sum_{i=1}^4 E_i^{1/2}\alpha_i+4\i \sum_{i=1}^2e^{\i\frac{\a_i-\theta_i+\i \rho+\i\rho_w}{2}}-4\i \sum_{i=3}^4e^{-\i\frac{\a_i-\theta_i-\i \rho-\i\rho_w}{2}} \bigg)+\text{suppressed\,}\bigg\rbrace\,,
\end{equation}
where the terms which we did not show would be at most of order $e^{-2 E^{1/2}T_i} N$ and so would \emph{not} affect the saddle-point equations. This effective action only depends on the combination
\begin{equation}
  \i \theta_1+\rho+\rho_w\,,\quad \i\theta_4-\rho-\rho_w
\end{equation}
and ditto for $\theta_2$ and $\theta_3$, respectively. This means that in this approximation $\rho$ and $\rho_w$ are not fixed, so that they remain as soft modes. The saddle for $\alpha_i$ remains at~\eqref{eq:cl-alpha_i}, and for every values of $\rho,\rho_w$ (the saddle-point manifold, or \emph{approximate} saddle-point manifold to be precise) the $\theta_i$ are pinned to
\begin{equation}
  e^{\i \frac{\theta_1}{2}}=e^{\i \frac{\a_1}{2}}\,\frac{E^{1/2}}{2}e^{-\frac{\rho+\rho_w}{2}}\,,\quad  e^{-\i \frac{\theta_4}{2}}=e^{-\i \frac{\a_4}{2}}\,\frac{E^{1/2}}{2}e^{-\frac{\rho+\rho_w}{2}}\,.
\end{equation}
Evaluating the original action~\eqref{eq:3.38ac} (including Boltzmann weights) on these configurations, one ends up with the soft mode action with soft modes $\rho$ and $\rho_w$ to be integrated over%
\footnote{The term $N 2\pi E^{1/2}$ denotes the partition function at fixed energy and should be stripped off as part of the normalization, akin to the $1/Z(\beta)$ in \eqref{eq:2point-JT}.}
\begin{equation}
  \exp \bigg\lbrace N 2\pi E^{1/2} + \frac{N}{Z} \bigg(\frac{E^{1/2}}{2}e^{\rho}-1\bigg)\bigg(\frac{E^{1/2}}{2}e^{\rho_w}-1\bigg)\bigg\rbrace\,,
\label{eq:softmodeaction}
\end{equation}
with the ``crossratio''~\cite{Lam:2018pvp,Maldacena:2016upp}
\begin{equation}
  \frac{1}{Z}=4 E^{1/2}\cos(\a)e^{-2 E^{1/2}\abs{T_1}}+4 E^{1/2}\cos(\a)e^{-2 E^{1/2}\abs{T_2}}\,.
\label{eq:crossratio}
\end{equation}
One can in fact check that this holds for all signs of $T_1$ and $T_2$.

Following~\cite{Maldacena:2016upp,Stanford:2021bhl}, it is convenient to introduce%
\footnote{From here on, we will work again with $N=1$ to be more consistent with our notation throughout the rest of the paper.}
\begin{equation}
  x=\frac{E^{1/2}}{2}e^{\rho}-1\,,\quad x_w=\frac{E^{1/2}}{2}e^{\rho_w}-1\,.
\label{eq:x-x_w}
\end{equation}
For opposite signs of $T_1$ and $T_2$, one finds
\begin{equation}
  e^{-\ell/2} = \frac{e^{-\ell_\text{bare}/2}}{x+1}\,,\quad e^{-\ell_w/2}=\frac{e^{-\ell_{w\,\text{bare}}/2}}{x_w+1}\,,
\label{eq:soft_l-l_w}
\end{equation}
with the bare lengths given by the classical saddle~\eqref{eq:4pt-lengths}. In the spirit of~\eqref{eq:A_Teff_l} one then writes
\begin{align}
  G_{\Delta\,\Delta_w\,\text{disk}}(T_{1},T_{2}) &= \int_{-\infty}^{+\infty}\d\ell\,\mathcal{A}_{\Delta_w\,\text{disk}}(T_{1},T_{2}, \ell)\,e^{-\Delta \ell}
\nonumber\\
  &=e^{-\Delta \ell_\text{bare}}e^{-\Delta_w \ell_{w\,\text{bare}}}\int\d x\,\d x_w \,e^{x x_w/Z}\,(x+1)^{-2\Delta}(x_w+1)^{-2\Delta_w}
\nonumber\\
  &=\frac{e^{-\Delta \ell_\text{bare}}e^{-\Delta_w \ell_{w\,\text{bare}}}}{\Gamma(2\Delta_w)}\int_0^\infty \d x\,x^{2\Delta_w-1}e^{-x}\,(xZ+1)^{-2\Delta}
\nonumber\\
  &=e^{-\Delta \ell_\text{bare}}e^{-\Delta_w \ell_{w\,\text{bare}}}\, _2{}F_0(2\Delta,2\Delta_w;;-Z)\,.
\label{eq:kummer}
\end{align}
This reproduces equation~(6.57) in~\cite{Maldacena:2016upp}, or (5.6)~and~(5.7) in~\cite{Lam:2018pvp}.%
\footnote{This hypergeometric function $_2{}F_0(\cdot,\cdot;;-Z)$ does not converge for all $Z$. It should be treated as the asymptotic series expansion of the Kummer U: $Z^{-2\Delta} U(2\Delta, 1-2\Delta-2\Delta_w,1/Z)$. This is the form presented in most literature (such as~\cite{Maldacena:2016upp,Lam:2018pvp}), but the symmetry between $\Delta$ and $\Delta_w$ is obscure in that form. The $_2{}F_0$ function in~\eqref{eq:kummer} highlights the symmetry.}
To go from the second line to the third line we use the Hankel contour and the definition of the reciprocal Gamma function; for details see~\cite{wiki}. The contours and measure are only correct starting from the third line. It would be interesting to track the contours through the whole calculation. In the probe approximation and for large $Z$, this expression simplifies to
\begin{equation}
  {G}_{\Delta\,\Delta_w\,\text{disk}}(T_{1},T_{2})=e^{-\Delta \ell_\text{bare}}e^{-\Delta_w \ell_{w\,\text{bare}}}\,Z^{-2\Delta}\,,\quad Z\gg 1\,,\quad T_1 T_2<0\,.
\label{eq:4pt-simplified}
\end{equation}
In the tau scaling limit $\abs{T_i}\to \infty$, we are always in the regime $Z\gg 1$ of this integral according to~\eqref{eq:crossratio}. This means that when the signs of $T_1$ and $T_2$ are opposite, a very strong shockwave interaction has been involved. Equation \eqref{eq:4pt-simplified} reproduces the result~\eqref{eq:l-disk-shock} of the classical shockwave calculation of~\cite{Stanford:2014jda}%
\footnote{This equation only concerns contributions which are exponentially large in $e^{\S}$.}
\begin{equation}
  \ell=2E^{1/2}(\abs{T_1}+\abs{T_2})\,,\quad T_1 T_2<0\,,
\end{equation}
It is reassuring to see this reappear from the exact JT gravity amplitude~\eqref{eq:A_disk}. Thus $T_1 T_2<1$ involves large crossratios $Z$ and is therefore preparing a dangerous slice.

For \emph{equal} signs of $T_1$ and $T_2$ one finds that $\ell$ is independent of $x$, with $x$ defined in~\eqref{eq:x-x_w}, whereas $\ell_w$ is identical to~\eqref{eq:soft_l-l_w}. In this case, the $x$ integral gives a delta function $\delta(x_\omega)$, such that both $\ell$ and $\ell_w$ localize onto their classical bare values~\eqref{eq:4pt-lengths}. In other words, one finds
\begin{equation}
  {G}_{\Delta\,\Delta_w\text{disk}}(T_{1},T_{2})=e^{-\Delta \ell_\text{bare}}e^{-\Delta_w \ell_{w\,\text{bare}}}\,,\quad T_1 T_2>0\,.
\label{3.54bis}
\end{equation}
This is, therefore, the case in which shockwaves might as well have been ignored. The geometry of the $\ell$ slice was not affected by the presence of the $\mathcal{O}_{\Delta_w}$ particles. Again, this matches the discussion of~\cite{Stanford:2014jda}, but now in the specific case of JT gravity.

After normalization (see footnote~\ref{fn:20}), equations~\eqref{eq:4pt-simplified} and \eqref{3.54bis} reproduce the claimed result~\eqref{eq:A-delta}
\begin{equation}
  \mathcal{A}_{\Delta_w\,\text{disk}}(T_{1},T_{2}, \ell)=\delta(\ell-2E^{1/2}(\abs{T_1}+\abs{T_2}))\,,
\end{equation}
together (as explained) with the claimed classification of safe and dangerous slices~\eqref{eq:regions-disk-eff}
\begin{align}
  T_1 T_2<0 \text{ or }\abs{T_{w}}&>\abs{T}\quad \text{dangerous}
\nonumber\\
  T_1 T_2>0 \text{ or }\abs{T_w}&<\abs{T}\quad \text{safe}
\label{eq:Teff-criterion}
\end{align}
%

\subsection{Conclusion}
\label{survival}

By now we have proven the steps leading up to equation~\eqref{eq:P_safe-T1-T2} in section~\ref{sect:log}. As explained there, the probabilities that the dual bulk slice is safe and dangerous, respectively, are computed as
\begin{align}
    P_\text{safe/danger}(T_1,T_2)=\int_{\text{safe/danger}}\! \d T_{1\,\text{eff}}\,\d T_{2\,\text{eff}}\,\mathcal{F}(T_{1\,\text{eff}}\rvert T_1)\mathcal{F}(T_{2\,\text{eff}}\rvert T_2)\,.
\label{eq:P_safe-danger}
\end{align}
As shown in~\eqref{eq:Teff-criterion}, the safe region corresponds to either both \emph{effective} times positive $T_{1\,\text{eff}}\,,T_{2\,\text{eff}}>0$ or both times negative. The dangerous region is where $T_{1\,\text{eff}}$ and $T_{2\,\text{eff}}$ have opposite signs. This simple criterion allows us to decompose~\eqref{eq:P_safe-danger} into the ``elementary'' probabilities~\eqref{eq:Pexp-Pcont}
\begin{equation}
  \boxed{P_\text{safe}(T_1,T_2)=P_\text{exp}(T_1)P_\text{exp}(T_2)+P_\text{cont}(T_1)P_\text{cont}(T_2)}\,,
\label{eq:P_safe-final}
\end{equation}
and similarly
\begin{equation}
  P_\text{danger}(T_1,T_2)=P_\text{exp}(T_1)P_\text{cont}(T_2)+P_\text{cont}(T_1)P_\text{exp}(T_2)\,.
\label{eq:danger-final}
\end{equation}
Here, the probabilities for negative times should be understood as
\begin{equation}
  P_\text{exp}(T_i<0) = P_\text{cont}(-T_i>0)\,,
\label{eq:P-negative}
\end{equation}
and vice versa. The probabilities $P_\text{exp}(T_i)$ and $P_\text{cont}(T_i)$ for $T_i > 0$ are given by~\eqref{eq:PP-1} and \eqref{eq:PP-2} with the ``constant'' subtracted, which we graphically reproduce here for reader's convenience
\begin{equation}
    \begin{tikzpicture}[baseline={([yshift=-.5ex]current bounding box.center)}, scale=0.7]
    \pgftext{\includegraphics[scale=1]{kyoto8.pdf}} at (0,0);
    \draw (0,-0.1) node {$1/2$};
    \draw (4.1,-0.7) node {$P_\text{cont}(T_i)$};
    \draw (-1.4,-0.7) node {$P_\text{exp}(T_i)$};
    \draw (0,1.8) node {$1$};
    \draw (-2.6,-2.8) node {$T_H$};
    \draw (2.9,-2.8) node {$T_H$};
  \end{tikzpicture}
\end{equation}
Recalling that
\begin{equation}
  T_1=T-T_w\,,\quad T_2=T+T_w\,,
\end{equation}
this graphic immediately leads to a main conclusion of this section:\ for late enough times, the chances of encountering and not encountering a firewall (according to our definition) are fifty-fifty
\begin{equation}
  \boxed{P_\text{safe}(T)=P_\text{danger}(T)=\frac{1}{2}}\,,\quad T>T_H+T_w\,.
\end{equation}
To be more precise, at exponentially late times, there is a fifty-fifty chance that the dual slice contains a strongly back-reacting shockwave in the states we study in this section.

Another interesting regime is $0<T_2 \ll T_H$, which corresponds to making a perturbation on the left, and having all the time evolution occur on the right. This might be closest in spirit to a setup obtained from collapse. One finds that the $T_2$ slice is expanding ($P_\text{exp}(T_2)=1$, $P_\text{cont}(T_2)=0$), and therefore
\begin{equation}
  P_\text{safe}(T_1,T_2)=P_\text{exp}(T_1)\,,\quad P_\text{danger}(T_1,T_2)=P_\text{cont}(T_1)\,.
\label{eq:collapsep}
\end{equation}
This is thus a specific setup in which all expanding slices are safe and all contracting slices are dangerous, which is what \cite{Stanford:2022fdt} had in mind. Again, the odds of finding expanding (safe) and contracting (dangerous) slices asymptote to fifty-fifty at exponentially late times.

Before proceeding, we tie up one loose end. In the discussion above, we have ignored contributions from the second diagram in~\eqref{eq:G_4pt-wormhole}. This represents the (physically very real) possibility that the matter particle created by the insertion of $\mathcal{O}_{\Delta_w}$ is carried away by a bra-ket wormhole in such a way that it bypasses (or avoids) the dual interior slice altogether. We call such events \emph{avoided crossings}. Avoided crossings most definitely correspond to safe interior slices (as there are no particles, highly boosted or otherwise, in the slice altogether). Thus, these would contribute to $P_\text{safe}(T)$ and, if dominant, might make the interior safe. However, as we show in appendix~\ref{app:avoided}, the opposite is actually true. As compared to the contributions discussed here, the avoided crossing is suppressed by a power $1/T_H^2\sim e^{-2\S}$ and should be ignored in the tau-scaling limit (which involves $T_H\to\infty$).

\subsection{Effective time-folds}
\label{sect:folds}

In light of the generalization in section~\ref{sec:multiple}, let us provide some additional intuition for the classification \eqref{eq:Teff-criterion} of safe/dangerous slices. We can think of the correlator~\eqref{eq:eucl4pt} as
\begin{equation}
  \bra{\Psi}\mathcal{O}_{\Delta\,\text{L}}\mathcal{O}_{\Delta\,\text{R}}\ket{\Psi}\,,
\label{eq:wavef-sq}
\end{equation}
where the state being probed is the following evolution of the TFD
\begin{equation}
  \ket{\Psi}=e^{-\i H_\text{L} T_2}\, \mathcal{O}_{\Delta_w\,\text{L}}\,e^{-\i H_\text{L} T_1}\ket{\Psi_\text{TFD}}\,.
\label{eq:timefolds}
\end{equation}
We used $H_\text{R}\ket{\Psi_\text{TFD}}=H_\text{L}\ket{\Psi_\text{TFD}}$ to put all Hamiltonian evolution on the left degrees of freedom. States of this type (and their dual) were analyzed in detail in~\cite{Stanford:2014jda}, who more generally considered
\begin{equation}
  \ket{\Psi}=e^{-\i H_\text{L} T_{n+1}} \prod_{i=1}^n \left( \mathcal{O}_{\Delta_{w_i}\,\text{L}}\,e^{-\i H_\text{L} T_i} \right) \ket{\Psi_\text{TFD}}\,.
\label{eq:foldgen}
\end{equation}
Depending on the relative signs of $T_1$ and $T_2$ in~\eqref{eq:timefolds} the boundary ``time-fold'' associated with~\eqref{eq:timefolds} is time-ordered (TO) or ``out-of-time-ordered'' (OTO)~\cite{Heemskerk:2012mn,Susskind:2013lpa}. Following notation of~\cite{Stanford:2014jda}, a fully time-ordered version of the time-fold preparing the state~\eqref{eq:foldgen} is pictured as
\begin{equation}
    \begin{tikzpicture}[baseline={([yshift=-.5ex]current bounding box.center)}, scale=0.7]
        \pgftext{\includegraphics[scale=1]{fire9.pdf}} at (0,0);
        \draw (-0.8,1.5) node {$T_{n+1}$};
        \draw (0,-0.58) node {$\dots$};
        \draw (-0.6,-2) node {$T_1$};
    \end{tikzpicture}
\end{equation}
whereas generic time ordering (signs of $T_i$) is pictured in~\cite{Stanford:2014jda} as a folded time contour
\begin{equation}
    \begin{tikzpicture}[baseline={([yshift=-.5ex]current bounding box.center)}, scale=0.7]
        \pgftext{\includegraphics[scale=1]{fire10.pdf}} at (0,0);
        \draw (1.1,-0.5) node {$T_{1}$};
        \draw (-1.4,0.2) node {$T_{n+1}$};
        \draw (0.1,-0.9) node {$\dots$};
    \end{tikzpicture}
\label{eq:folded-pic}
\end{equation}

The main focus of~\cite{Stanford:2014jda} was to discuss the complexity of these states or correspondingly~\cite{Susskind:2014rva} the length $\ell$ of the dual bulk slice (focusing now on 2d gravity). The authors considered times much shorter than $T_H$, so wormholes did not play any role, and pictures such as~\eqref{eq:folded-pic} in this case represent physics occurring in the bulk at the classical level. We have seen, however, that wormholes change this. In particular, one can represent~\eqref{eq:ampbb} and \eqref{eq:4pt-exp} using \emph{effective time-folds}%
\footnote{As a reminder, we only draw the Lorentzian time-fold for a ket. To compute the actual expectation value, we would need another copy of time-fold for the bra (along with the Euclidean circles). See, for example, equation~(1.12) in~\cite{Stanford:2020wkf} for a more complete drawing.}
\begin{equation}
    \bra{\Psi}\dots \ket{\Psi}=\begin{tikzpicture}[baseline={([yshift=-.5ex]current bounding box.center)}, scale=0.7]
        \pgftext{\includegraphics[scale=1]{fire7.pdf}} at (0,0);
        \draw (-0.6,0.6) node {$T_2$};
        \draw (-0.6,-1.2) node {$T_1$};
    \end{tikzpicture}\quad+\text{ wormholes }=\int_{-\infty}^{+\infty}\d T_{1\,\text{eff}}\,\mathcal{F}(T_{1\,\text{eff}}\rvert T_1)\int_{-\infty}^{+\infty}\d T_{2\,\text{eff}}\,\mathcal{F}(T_{2\,\text{eff}}\rvert T_2)\quad
    \begin{tikzpicture}[baseline={([yshift=-.5ex]current bounding box.center)}, scale=0.7]
        \pgftext{\includegraphics[scale=1]{fire8.pdf}} at (0,0);
        \draw (0.9,-0.9) node {$T_{1\,\text{eff}}$};
        \draw (-0.95,0.2) node {$T_{2\,\text{eff}}$};
    \end{tikzpicture}
\label{eq:pic-eff-timefold}
\end{equation}
where we now take each contour to represent a \emph{unique} bulk slice arising effectively through wormhole contributions. We emphasize that these contours do \emph{not} represent boundary conditions in this equation! They are mnemonics that capture classical (effective) bulk slices.

In the language developed above, the criterion~\eqref{eq:regions-disk-eff} becomes (perhaps) more intuitive
\begin{align}
  \boxed{\begin{aligned}\text{effective OTO}\quad&\text{ dangerous}\\\text{effective TO}\quad& \text{ safe}\end{aligned}}\,\,.
\end{align}
By ``OTO'' we mean that there is (at least) one switchback in the \emph{effective} time contour that represents the dual semiclassical bulk slice. A ``switchback''~\cite{Stanford:2014jda} is a fold in the time-fold. Thus, one way to phrase our findings is that wormholes may replace TO time-folds as boundary conditions with \emph{effective} OTO time-folds. This results in dangerous shockwaves in unexpected places.

\subsection*{Warning about nomenclature}

We warn the reader that sometimes in the literature other criteria have been used to refer to a correlator as TO or OTO; here we have followed~\cite{Stanford:2014jda}. In particular, oftentimes one refers to the four-point function we considered in \eqref{eq:eucl4pt} as OTOC for generic (complex) times between all operators, simply because the operators are ordered as $\mathcal{O}_\Delta \mathcal{O}_{\Delta\,w}\mathcal{O}_\Delta \mathcal{O}_{\Delta\,w}$ along the boundary. This is \emph{not} the nomenclature we follow.

\section{Exponentially dangerous states}
\label{sec:multiple}

In this section, we consider the logical generalization of section~\ref{sec:one}. What if we consider states prepared with multiple early perturbations? How dangerous are those?

We consider states obtained by perturbing the TFD on the left boundary at multiple times $t = -T_{w_i}$ ($i = 1,\ldots,n$), where we take
\begin{equation}
  T_{w_1} < \ldots < T_{w_n}.
\end{equation}
One could be interested in the case in which these perturbations are distributed in a ``typical'' manner. There are many ways to define typicality, as typicality refers to a choice of ensemble. In our setup, the most natural ensemble may be to consider early perturbations at times taking values on the whole real axis (within the recursion time). Then the vast majority of configurations will have (much) more than the Heisenberg time of separation between each particle. We show that such states almost always have firewalls, even though classically (ignoring wormholes) they would appear to be perfectly safe. 

We do not want to argue that this ensemble is physically very relevant (which would mean it would be a good representative for black holes formed from gravitational collapse). Instead, we want to point out just quite \emph{how} dramatic the situation can get once you properly account for wormholes.

\subsection{Multiple shocks}

We again consider the amplitude~\eqref{eq:wavef-sq}, but now it involved the state~\eqref{eq:foldgen} with $n$ perturbations
\begin{equation}
  \ket{\Psi}=e^{-\i H_\text{L} T_{n+1}}\prod_{i=1}^n \left(\mathcal{O}_{\Delta_{w_i}\,\text{L}}\,e^{-\i H_\text{L} T_i}\right)\ket{\Psi_\text{TFD}}\,,
\end{equation}
where
\begin{equation}
  T_1 = T - T_{w_n},
\qquad
  T_i = T_{w_{n+2-i}} - T_{w_{n+1-i}}\text{ for } i = 2\dots n\,,
\qquad
  T_{n+1} = T_{w_1} + T.
\label{eq:T_i-s}
\end{equation}
The dual semiclassical geometry (ignoring wormholes) was discussed in~\cite{Stanford:2014jda}. Applied to 2d gravity, and for $T_i$'s much larger than the scrambling time, one finds a slice with $n$ shockwaves and total length
\begin{equation}
  \ell=2 E^{1/2}\sum_{i=1}^{n+1}\abs{T_i}\,.
\label{eq:T_i-length}
\end{equation}
Equations~\eqref{eq:T_i-s} and \eqref{eq:T_i-length} are the generalization of~\eqref{eq:l-disk-shock}. The length~\eqref{eq:T_i-length} should be compared to the length of the dual slice in the \emph{absence} of the shockwaves (or particle insertions)
\begin{equation}
  \ell_\text{bare}=2 E^{1/2}\abs{T_\text{total}}\,,\quad T_\text{total}=\sum_{i=1}^{n+1}T_i=2T\,.
\end{equation}
This implies that significant backreaction occurs as soon as \emph{not all} the $T_i$'s have identical signs, or in other words as soon as \emph{at least} one switchback has occurred in the notation of~\eqref{eq:folded-pic}. The intuitive criterion~\eqref{eq:regions-eff-tf} is thus the correct criterion in this more general case as well, in determining whether or not a bulk slice is dangerous or safe ($\sim$ has a strong shock or not).
We emphasize that time ordered (TO) in this context means that \emph{all signs} are identical, i.e.\ $\text{sgn}(T_1)=\dots =\text{sgn}(T_{n+1})$.

We now claim that the precise generalization of~\eqref{eq:pic-eff-timefold} in the tau-scaling limit is
\begin{equation}
    \bra{\Psi}\dots \ket{\Psi}=\begin{tikzpicture}[baseline={([yshift=-.5ex]current bounding box.center)}, scale=0.7]
        \pgftext{\includegraphics[scale=1]{fire9.pdf}} at (0,0);
        \draw (-0.8,1.5) node {$T_{n+1}$};
        \draw (0,-0.58) node {$\dots$};
        \draw (-0.6,-2) node {$T_1$};
    \end{tikzpicture}\quad+\text{ wormholes }=\prod_{i=1}^{n+1}\int_{-\infty}^{+\infty}\!\d T_{i\,\text{eff}}\,\mathcal{F}(T_{i\,\text{eff}}\rvert T_i)\,\,\,
    \begin{tikzpicture}[baseline={([yshift=-.5ex]current bounding box.center)}, scale=0.7]
        \pgftext{\includegraphics[scale=1]{fire10.pdf}} at (0,0);
        \draw (1.3,-0.5) node {$T_{1\,\text{eff}}$};
        \draw (-1.6,0.2) node {$T_{n+1\,\text{eff}}$};
        \draw (0.1,-0.9) node {$\dots$};
    \end{tikzpicture}\,\,.
\label{eq:gen-timefold}
\end{equation}
Here, we consider the generalized tau-scaling limit
\begin{equation}
  T_1,\,T_2,\,\ldots,\,T_{n+1},\,e^{\S}\to \infty \,,\qquad T_1 e^{-\S},\, T_2 e^{-\S},\, \ldots,\, T_{n+1} e^{-\S}\text{ fixed}.
\label{eq:gen-tauscaling}
\end{equation}
The derivation of \eqref{eq:gen-timefold} consists of several steps which are quite identical to those in section~\ref{sec:one}. Leaving the details to the interested reader, we simply summarize the main steps
\begin{enumerate}
\item
In the generalization of~\eqref{eq:eucl4pt}, let us label the energies bordering boundary segments with Lorentzian time $\pm T_i$ as $E_{i\,\text{bra}}$ and $E_{i\,\text{ket}}$. Such regions are on opposite sides of the interior slice (red). Particle lines (blue) separate energies with different indices $i$. All regions get a Euclidean regulator. Then in our regime of interest the appropriate integration kernel (generalizing~\eqref{eq:rho-factor}) is
\begin{equation}
  \frac{\rho(E_1\dots E_{2n+2})}{\prod_{i=1}^{2n+2} \rho_0(E_i)}=\prod_{i=1}^{n+1}\bigg(1+\frac{\delta(\omega_i)}{\rho(E)}-\frac{\sin(\pi\rho(E)\omega_i)^2}{\pi\rho(E)^2\omega_i^2}\bigg)\,,\quad \omega_i=E_{i\,\text{bra}}-E_{i\,\text{ket}}\,.
\label{eq:gen-kernel}
\end{equation}
The reason for this specific replacement is the Boltzmann weight $\omega_i T_i$, which at tau-scaling times favors bra and ket energies to be exponentially close. This forces us to consider bra-ket wormholes within each index $i$, such as those drawn in~\eqref{eq:eucl4pt}. On the other hand, there is no sense in which the integral prefers energies with different indices to be \emph{exponentially} close together, since their energy differences have finite Euclidean Boltzmann weights (the aforementioned regulators). Therefore, other wormholes than those counted in~\eqref{eq:gen-kernel} contribute negligibly.
\item
With this result, the generalization of~\eqref{eq:G_T1T2-decomp} is proven. Indeed, the convolution theorem can just be applied again. This suffices to prove~\eqref{eq:gen-timefold}. We would now like to go one step further and prove the generalization of~\eqref{eq:P_safe-danger}. This is true \emph{if} the disk amplitude decomposes similarly to~\eqref{eq:A_Teff_l}
\begin{equation}
  G_{\Delta\,\Delta_{w_1}\dots \Delta_{w_n}\,\text{disk}}(T_{1},\ldots,T_{n+1})=\int_{-\infty}^{+\infty}\!\d\ell\,\mathcal{A}_{\Delta_{w_1}\dots \Delta_{w_n}\,\text{disk}}(T_{1},\ldots,T_{n+1}, \ell)\,e^{-\Delta \ell}\,,
\end{equation}
with (in the semiclassical limit and after normalization)
\begin{equation}
  \mathcal{A}_{\Delta_{w_1}\dots \Delta_{w_n}\,\text{disk}}(T_{1},\ldots,T_{n+1}, \ell)=\delta(\ell-\ell(T_i))\,,\quad \ell(T_i)=2 E^{1/2}\sum_{i=1}^{n+1}\abs{T_i}\,.
\label{eq:410}
\end{equation}
We did not check this explicitly for the multiple-shocks setup, but it is largely obvious. Equation~\eqref{eq:410} is identical to the statement that the two-point function in the shockwave geometry is classically $e^{-\Delta \ell(T_i)}$.%
\footnote{We normalize the amplitude with that for fixed boundary times, not effective times. One might worry that dependence on effective times might creep in via the factor $e^{-\Delta_i \ell_{w_i}(T_{j\,\text{eff}})}$. This does not happen. Indeed, the perturbation particles on-shell follow geodesics in an unperturbed TFD at $t=0$, because the ``total time'' left and right of them adds up to zero, and therefore the length $\ell_{w_i}$ is in fact independent of time (and hence also of effective times). So there is no effective time dependence coming in via the normalization, at least classically. This statement is the generalization of footnote~\ref{fn:20}.}
The least obvious part is the reasonable claim that the exact Schwarzian disk amplitude is dominated by that ``classical'' shockwave geometry. In section~\ref{sect:3.3}, we checked carefully that this is indeed the case, albeit only for the setup with one shock $n=1$.
\end{enumerate}
This results in the following generalization of~\eqref{eq:P_safe-final}
\begin{equation}
  \boxed{P_\text{safe}(T_1,\ldots,T_{n+1})=P_\text{exp}(T_1)\dots P_\text{exp}(T_{n+1})+P_\text{cont}(T_1)\dots P_\text{cont}(T_{n+1})}\,.
\label{eq:multiple-P_safe}
\end{equation}
This counts only those contributions where \emph{all} signs of $T_{i\,\text{eff}}$ are equal. Any other combination of signs has at least one switchback, and should be considered dangerous
\begin{equation}
  P_\text{danger}(T_1,\ldots,T_{n+1})=\sum_{\text{signs unequal}}P_\text{exp}(\pm T_1)\dots P_\text{exp}(\pm T_{n+1})\,,
\label{eq:multiple-P_danger}
\end{equation}
with probabilities at negative times computed as in~\eqref{eq:P-negative}. We will now analyze these probabilities in various scenarios, depending on the parametric choices of boundary times $T_i$.

\subsection{Firewall probabilities}
The clearest physical picture is when one considers either $T_i\ll T_H$ or $T_i\gg T_H$. In the former case, one recovers classical physics, the geometry simply expands
\begin{equation}
  P_\text{exp}(T_i)=1\,,\quad T_i\ll T_H\,.
\label{eq:pure-exp}
\end{equation}
In the latter case, we probe the plateau region where the chances of expanding and contracting branches are equal
\begin{equation}
  P_\text{exp}(T_i)=\frac{1}{2}\,,\quad T_i\gg T_H\,.
\label{eq:pplat}
\end{equation}
To demonstrate our methods, we will explore three different scenarios.
\begin{enumerate}
\item
The $n$ particles are all separated by $T_i\gg T_H$. In this case, all the factors in~\eqref{eq:multiple-P_safe} have reached their probability plateaus~\eqref{eq:pplat}. Thus, for extremely late times $T>T_{w_n}+T_H$, we have
\begin{equation}
  P_\text{safe}(T)=\frac{2}{2^{n+1}}=\frac{1}{2^{n}}\,.
\end{equation}
Similarly, by counting all the dangerous permutations in~\eqref{eq:multiple-P_danger} we obtain
\begin{equation}
  \boxed{P_\text{danger}(T)=1-\frac{1}{2^{n}}}\,,\quad T>T_{w_n}+T_H\,.
\end{equation}
This is the result announced in the introduction~\eqref{eq:intro-firewall}, and one of the main points of our paper. For late times, these states (which as pointed out at the beginning of this section, are typical in some sense) \emph{almost certainly} have firewalls if we consider many perturbations $n$.
\item
The particles are bunched together around $t=0$ at timescales much shorter than the Heisenberg time, but longer than the scrambling time $T_H \gg T_2,\ldots,T_n \gg T_\text{S}=(\b/2\pi) \log \S$. In this case, the $n-1$ intermediate times give purely expanding probabilities~\eqref{eq:pure-exp}. This results in
\begin{equation}
  P_\text{safe}(T_1,\ldots,T_{n+1})=P_\text{exp}(T_1)P_\text{exp}(T_{n+1})
\end{equation}
and
\begin{equation}
  P_\text{danger}(T_1,\ldots,T_{n+1})=P_\text{cont}(T_1)P_\text{cont}(T_{n+1})+P_\text{exp}(T_1)P_\text{cont}(T_{n+1})+P_\text{cont}(T_1)P_\text{exp}(T_{n+1})\,.
\end{equation}
This results for $T>T_H$ in the plateau
\begin{equation}
  \boxed{P_\text{danger}(T)=\frac{3}{4}}\,,\quad T>T_H\,.
\end{equation}
It may look puzzling why the contracting-contracting term is dangerous if we compare with the $n=1$ case in section~\ref{sec:one} (see~\eqref{eq:danger-final}). The reason is that the ordering of the $n>1$ operator insertions already picks a ``preferred'' time axis. Indeed, the contracting-contracting case has a non-trivial time-fold~\cite{Stanford:2014jda}:
\begin{equation}
        \begin{tikzpicture}[baseline={([yshift=-.5ex]current bounding box.center)}, scale=0.7]
        \pgftext{\includegraphics[scale=1]{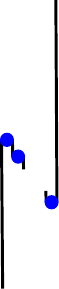}} at (0,0);
        \draw (1,1) node {$T_{1}$};
        \draw (0,-0.58) node {$\dots$};
        \draw (-1.3,-1.5) node {$T_{n+1}$};
    \end{tikzpicture}
\label{4.19}
\end{equation}
Thus the dual geometry still contains a shock. While this shock is ``less severe'' than in the other cases, it \emph{is} dangerous. Indeed, the effect on the geodesic $\ell$, whilst much less than $e^{\S}$, is still much larger than $\log \S$.

If the perturbations are bunched around $T_w\gg T_H$, we reach the same asymptotics for $T>T_H+T_w$.
\item
If successive particles are separated by timescales less than the scrambling time (such as thermal) one may essentially treat them as one particle in the switch-back diagrams of~\cite{Stanford:2014jda}. Therefore, in a scenario where all the particles are bunched around $t=0$ and $T_2,\ldots,T_n \ll T_S$, one finds
\begin{equation}
  \boxed{P_\text{danger}(T)=\frac{1}{2}}\,,\quad T>T_H\,.
\end{equation}
Indeed, in this case the contracting-contracting configuration~\eqref{4.19} has essentially no switch-back; in other words, the shockwave does not severely backreact on $\ell$~\cite{Stanford:2014jda}. Therefore, in this case the contracting-contracting term should be considered to contribute to $P_\text{safe}(T)$.
\end{enumerate}

In summary, black holes created with early perturbations on the other side of the TFD are generically very dangerous at post-Heisenberg times. Depending on the detailed setup and one's notion of typicality, at the very best the probability of a firewall is $1/2$. At worst, firewalls are \emph{guaranteed}.

\section{Concluding remarks}
\label{sec:concl}

To end this paper, we will point out that our results (using as an example~\eqref{eq:F-Fourier}) can be obtained by summing (following~\cite{Saad:2022kfe,Blommaert:2022lbh,Weber:2022sov}) a perturbatively convergent series (in genus) of wormhole amplitudes, and we will also identify the corresponding Lorentzian wormhole geometries~\cite{Blommaert:2023vbz,Usatyuk:2022afj}. This is discussed, respectively, in sections~\ref{sect:plat} and \ref{sect:disclor}.

Before doing so, however, we propose several ways to potentially improve on our setup. Our general attitude is that despite some shortcomings in our setup, our results seem physically insightful. We think that our techniques and ideas will also help in attacking the improvements that we propose below.

\subsection{Room for improvement}
\label{subsec:improv}

Here we list potential points of critique on our setup and how to improve on them. 
\begin{enumerate}
\item
We are not describing measurements performed by an infalling observer in quantum gravity, as it is not known how to describe such infalling observers in quantum mechanics. (For some interesting recent progress on this problem see for instance~\cite{Chandrasekaran:2022cip,Leutheusser:2021qhd}.) It might be that our physical conclusion changes a lot when an observer is included. As a very crude approximation, one could contemplate for instance modeling an observer by a particle with $\Delta\to 0$ that stretches the interior slice.  More precisely, if we take the $\Delta\rightarrow 0$ limit while keeping $\Delta\gg1/T_H\gg 1/\Lambda$ with $\Lambda$ an IR cutoff for $T_\text{eff}$, then the exponential suppression $e^{-\Delta \ell(T_\text{eff})}$ ensures essentially that wormhole corrections in~\eqref{eq:F-Fourier} do not contribute
\begin{align}
  P_{\text{cont}\,\Delta}(T) &= \lim_{\Delta\rightarrow 0} \int^0_{-\Lambda} dT_\text{eff}\, \mathcal{F}(T_\text{eff}\rvert T) e^{-2\Delta \sqrt{E}\abs{T_\text{eff}}}=0\,,\quad P_{\text{exp}\,\Delta}(T)=1\,.
\end{align}
Here we considered $\Delta\ll 1/T$ such that $e^{-2\Delta\sqrt{E}T}\rightarrow 1$. This is not a realistic model; but it shows the potentially far-reaching consequences of carefully defining the observer.
\item
We considered probe matter perturbations, \emph{not} dynamical QFT with particle-antiparticle pairs. A consequence of this is that in our setup and that of~\cite{Stanford:2022fdt}, the pure TFD is \emph{obviously} safe. However, when particle-antiparticle pairs can be created, it is fathomable that wormholes could produce dangerous shocks in unexpected places. We believe such effects would be closer in spirit with the original firewall ideas~\cite{Almheiri:2012rt,Almheiri:2013hfa,Marolf:2013dba}. For instance, one might imagine (schematically) a surprising shock even in the pure TFD, due to a process like:
\begin{equation}
        \begin{tikzpicture}[baseline={([yshift=-.5ex]current bounding box.center)}, scale=0.7]
        \pgftext{\includegraphics[scale=1]{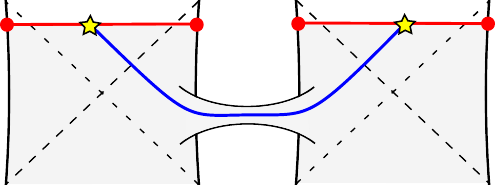}} at (0,0);
        \draw (-5,1.1) node {$\color{red}T_\text{eff}$};
        \draw (-5,-0.5) node {$\color{blue}T_{w\,\text{eff}}$};
\end{tikzpicture}
\end{equation}
Unfortunately studying dynamical matter on wormhole geometries is challenging. Indeed, in JT gravity it leads to UV divergences. To make further progress on this question, it seems one would have to first resolve that issue. One could start with studying one particle loop. Alternatively, one could consider a UV regulated q-deformation of JT gravity~\cite{Jafferis:2022wez}, which has a bulk interpretation as a different simple 2d dilaton gravity~\cite{Blommaert:2023wad,wip}. In that theory matter loops seem better behaved. Supersymmetric JT may also deal with these UV divergences \cite{Belaey:2023jtr}.
\item
As discussed in section~\ref{sect:gray}, we renormalized our amplitudes by subtracting off an infinite constant. We believe this is physically well motivated, following identical logic in~\cite{Iliesiu:2021ari}. Nevertheless, this is a subtle point, one that for instance Stanford and Yang~\cite{Stanford:2022fdt} put a lot of effort in trying to avoid.%
\footnote{We are not sure that factorizing the empty partition function as Stanford and Yang do is the best setup to improve on this. Physical observations involve measurements (as in our setup), while the empty partition function does not.}
Obviously in a UV complete theory such subtractions should not be required. From this point of view, it may be worthwhile trying to reformulate these types of questions in the matrix integral dual~\cite{Saad:2019lba} of JT gravity, and in particular in one member of the ensemble~\cite{Blommaert:2019wfy,Marolf:2020xie,Blommaert:2021gha,Blommaert:2021fob,Blommaert:2022ucs,Saad:2021rcu} (even though individual members of the ensemble may be dangerous for very different reasons~\cite{Kruthoff:2024gxc}). Progress on this front is being made~\cite{wipluca}.
\item
Orthogonal to the previous problems, it would be an improvement to mimic our setup for black holes in the sky (formed from a gravitational collapse), which are not a TFD with certain amount of perturbations on the left (which we studied). A first step in this direction would be to consider pure states in AdS. This should be possible, probably using end-of-the-world (EOW) branes in JT gravity~\cite{Kourkoulou:2017zaj,JafferisEOW}. What is the physically meaningful question to ask in an EOW brane setup? Naively, the slice is always dangerous when it ends on the EOW brane (spacetime ending is quite dramatic), but this conclusion is likely overly simplistic. A second step is to allow the black hole to evaporate, perhaps along the lines of~\cite{Penington:2019kki}. More realistic models suffer often from the lack of a simple exact Euclidean path integral description, and seem more difficult to study using our techniques.%
\footnote{See for instance~\cite{Nomura:2018kia,Nomura:2019qps,Nomura:2019dlz,Langhoff:2020jqa,Nomura:2020ska,Murdia:2022giv} for more discussion on black holes formed from collapse.}
\end{enumerate}

\subsection{Perturbative firewall probability plateaus}
\label{sect:plat}

We remind the reader of our semiclassical answer for the transition kernel~\eqref{eq:F-final}
\begin{equation}
  \mathcal{F}(T_\text{eff}\rvert T)=\frac{1}{2\pi \rho(E)^2}\int_{-\infty}^{+\infty}\!\d\omega\,e^{\i \omega (T_\text{eff}-T)}\bigg(\rho(E)^2+\delta(\omega)\rho(E)-\frac{\sin(\pi\rho(E)\omega)^2}{\pi\omega^2}\bigg)\,.
\end{equation}
This can be rewritten as
\begin{align}
  \mathcal{F}_\text{un-norm}(T_\text{eff}\rvert T)&=\delta(T-T_\text{eff})\, \rho(E)
\nonumber\\
  &\qquad +\frac{1}{2\pi\rho (E)}\,\frac{1}{\i \pi}\int_{-\i\infty}^{+\i\infty}\!\d\beta\,e^{2\beta E}\,Z_\text{conn}(\beta+\i(T-T_\text{eff}),\beta-\i(T-T_\text{eff}))\,.
\label{eq:F_un-norm}
\end{align}
Here, we have multiplied $\mathcal{F}(T_\text{eff}\rvert T)$ by $\rho(E)$, as we want to compare it with an un-normalized sum over geometries in the gravitational path integral. Furthermore we introduced
\begin{equation}
  Z_\text{conn}(\beta+iT,\beta-iT)= \int_0^{\infty}\! \d E\, e^{-2\beta E} \text{min} (|T|/2\pi,\rho(E))\,.
\label{eq:Z_conn-dom}
\end{equation}
One of the main points of~\cite{Saad:2022kfe,Blommaert:2022lbh,Weber:2022sov} was that this equation for the connected two-boundary amplitude is \emph{exact} for dilaton gravity (in the tau-scaling limit), and that it admits a Taylor series in $T$
\begin{equation}
  Z_\text{conn}(\beta+iT,\beta-i T)=\frac{T}{4\pi \beta}+\sum_{g=1}^{\infty} P_{g-1}(\beta)\, T^{2g+1}\,,
\label{eq:Z_conn}
\end{equation}
where the polynomial $P_{g-1}(\beta)$ is computed by a contour integral around the real axis
\begin{equation}
  P_{g-1}(\beta)=-\frac{1}{(2\pi)^{2g+1}(2g)(2g+1)}\oint_R\d E \,\rho(E)^{-2g}\,e^{-2\beta E}\,.
\end{equation}
Because $\rho(E)\sim e^{\S}$, this Taylor series is actually the gravitational genus expansion, with $g$ the number of handles (or wormholes). This series is reproduced by the Weil-Peterson polynomials in~\eqref{eq:rho_conn}~\cite{Saad:2022kfe,Blommaert:2022lbh,Weber:2022sov}. So, the sine kernel~\eqref{eq:sinekernel} of random matrix theory in the time domain~\eqref{eq:Z_conn-dom} is \emph{perturbatively} (in $g$) accessible in gravity (in the tau-scaling limit).%
\footnote{Backing off from this tau-scaling limit, one does have to consider non-perturbative effects~\cite{Saad:2019lba,Post:2022dfi,Altland:2020ccq,Eynard:2023qdr,Griguolo:2023jyy,Okuyama:2021eju}.}

Here we want to point out that the same is true for our un-normalized probability~\eqref{eq:F_un-norm}. In particular, at fixed temperature and following the same steps resulting in equation~(A.14) or (A.19) in~\cite{Blommaert:2023vbz}, one obtains the genus expansion%
\footnote{One can use this to compute the two-point function~\eqref{eq:2point-JT} in the tau scaling limit by inserting $e^{-\Delta \ell(T_\text{eff})}$ (taken from equation~\eqref{eq:ell}). However, in the tau-scaling limit this exponential backreacts heavily, since $\ell$ is exponentially large in the entropy. This thus projects onto $T_\text{eff}=0$. The equation for the two-point function then reduces to equation~(4.5) in~\cite{Blommaert:2023vbz}.}
\begin{align}
  \mathcal{F}_\text{un-norm}(T_\text{eff}\rvert T)&=\delta(T-T_\text{eff})\, Z(\beta)+\frac{1}{4\pi^2}\int_0^\infty\! \d E\,e^{-\beta E}\,\rho(E)^{-1}\,\abs{T-T_\text{eff}}\nonumber\\&\qquad-\,\sum_{g=1}^\infty \frac{1}{2g(2g+1)(2\pi)^{2g+2}}\oint_R\d E\,e^{-\beta E}\,\rho(E)^{-1-2g}\,\abs{T-T_\text{eff}}^{2g+1}\,.
\label{eq:gexpexp}
\end{align}
This sum is convergent~\cite{Saad:2022kfe,Blommaert:2022lbh}, so the firewall probability plateau in~\eqref{1.6intro} is perturbatively accessible.

\subsection{Lorentzian spacetimes}
\label{sect:disclor}

We now wonder how to obtain~\eqref{eq:gexpexp} from purely Lorentzian geometries. Besides an independent interest in Lorentzian wormhole geometries, this is relevant to us as it provides another indication that effective times $T_{i\,\text{eff}}$ determine the true Lorentzian slice of spacetime that is being probed. This interpretation was motivated more in sections~\ref{sect:2.1} and \ref{sect:semiwave}.

The relevant Lorentzian spacetimes are a mild modification of those discussed in section~4 of~\cite{Blommaert:2023vbz}, which we follow closely, and to which we refer readers for a more pedagogical explanation. Like in~\cite{Blommaert:2023vbz}, we do not have enough control of Lorentzian JT gravity to reproduce the full details of~\eqref{eq:gexpexp}. Instead, we \emph{can} reproduce the semiclassical ($\sim$ large energy) approximation%
\footnote{One could say that this is a poor approximation to~\eqref{eq:gexpexp}, which obtains its main contributions from very low energies \cite{Saad:2022kfe}. Classical physics is a poor approximation for low energies. Thus, semiclassically, the best one could hope for might be~\eqref{eq:prediction}.}
\begin{align}
  \mathcal{F}_\text{un-norm}(T_\text{eff}\rvert T)&\sim \int_{\Lambda_g}^\infty\d E\,e^{-\beta E}\,\rho(E)^{-1-2g}\,\abs{T-T_\text{eff}}^{2g+1}\,.
\label{eq:prediction}
\end{align}
This is reproduced, for $T_\text{eff}>0$ (expanding) and $g=0$, by the following geometries 
\begin{equation}
    \begin{tikzpicture}[baseline={([yshift=-.5ex]current bounding box.center)}, scale=0.7]
 \pgftext{\includegraphics[scale=1]{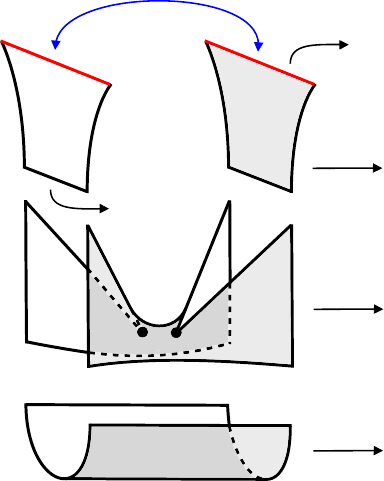}} at (0,0);
    \draw (4.7,1.25) node {expanding};
    \draw (4.9,-1.1) node {double cone};
    \draw (4.2,-3.6) node {gutter};
    \draw (-3.6,2.2) node {\color{red}$T_\text{eff}$};
    \draw (-4.1,-0.7) node {\color{red}$T-T_\text{eff}$};
    \draw (-0.6,0.5) node {$t=0$};
    \draw (3.7,3.3) node {$t=T_\text{eff}$};
    \draw (-0.7, 4.5) node {\color{blue}identify};
  \end{tikzpicture}
\label{eq:figexp}
\end{equation}
whereas for $T_\text{eff}<0$ (contracting) and $g=0$, the relevant (mostly) Lorentzian spacetimes are
\begin{equation}
    \begin{tikzpicture}[baseline={([yshift=-.5ex]current bounding box.center)}, scale=0.7]
 \pgftext{\includegraphics[scale=1]{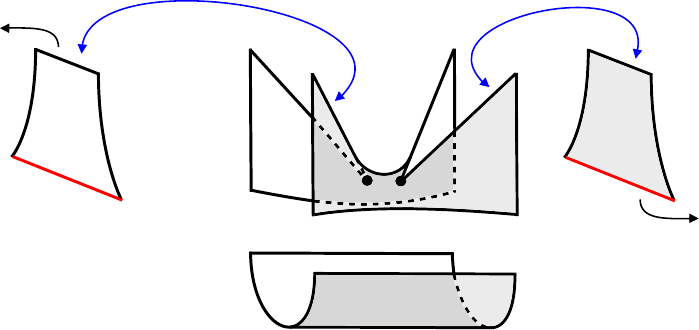}} at (0,0);
    \draw (7.5,-0.95) node {$t=T_\text{eff}<0$};
    \draw (-6.2,1) node {\color{red}$T_\text{eff}$};
    \draw (-2.9,0.5) node {\color{red}$T-T_\text{eff}$};
    \draw (-6.7,2.3) node {$t=0$};
    \draw (-2.5, 3.3) node {\color{blue}identify};
  \end{tikzpicture}
\label{eq:figcon}
\end{equation}
These are to be compared with equation~(4.16) in~\cite{Blommaert:2023vbz}. The ``gutter'' is half of the Euclidean wormhole, which is glued onto patches of the double-cone spacetime~\cite{Saad:2018bqo} in the way indicated (see (4.15) in~\cite{Blommaert:2023vbz} for more details). The black dots denote specific curvature singularity called ``crotches''~\cite{Blommaert:2023vbz,Louko:1995jw}. The point is that part of the time evolution imposed by the boundary conditions can be ``absorbed'' by a portion of double-cone spacetime. Any time slice of that double cone is identical to the global TFD at $t=0$, with metric and dilaton
\begin{equation}
  \d s=\d \rho=\frac{\d\s}{\sin(\s)}\,,\quad \Phi=E^{1/2}\cosh(\rho)=\frac{E^{1/2}}{\sin(\s)}\,,
\end{equation}
which means we can glue this smoothly to a $t=0$ TFD, and use the remainder of the boundary time evolution to expand that slice either into the future or past (depending on whether this remaining time $T_\text{eff}$ is positive or negative). For $T-T_\text{eff}<0$ one simply lets the double-cone evolve to the past.

A factor $\abs{T-T_\text{eff}}$ comes from the twist zero mode of the double cone~\cite{Blommaert:2023vbz}. For $g>0$ the spacetimes are those in~\eqref{eq:figexp} and \eqref{eq:figcon} with additional crotch singularities inserted at mirrored locations on the double-cone pieces of the $g=0$ spacetimes, in analogy to equation (4.18) in~\cite{Blommaert:2023vbz}. Their time coordinates on the double cone are zero modes and explain the additional powers of $\abs{T-T_\text{eff}}$ in~\eqref{eq:prediction}.%
\footnote{The saddle-point equations localize the additional crotches on the double cone piece. Indeed, there is no saddle on the expanding piece. This explains the powers of $\abs{T-T_\text{eff}}$ as opposed to simple powers of $T$.}
The power of $\rho(E)$ comes from the on-shell actions of the crotches~\cite{Blommaert:2023vbz}. So, summing over classical Lorentzian wormhole geometries, following the rules put forward in~\cite{Blommaert:2023vbz}, reproduces~\eqref{eq:prediction}. And indeed, the slice in which the measurement takes place has the geometry of the usual TFD, but at \emph{effective} time $t=T_\text{eff}$.

\section*{Acknowledgments}

We thank Jorrit Kruthoff, Geoff Penington, Douglas Stanford, Mykhaylo Usatyuk, Shunyu Yao, and Ying Zhao for useful discussions.
AB was funded by ERC-COG Grant NP-QFT No.\ 864583 and by INFN Iniziativa Specifica GAST.
The work of CHC and YN was supported in part by the Department of Energy, Office of Science, Office of High Energy Physics under QuantISED award DE-SC0019380 and contract DE-AC02-05CH11231.
The work of CHC was also supported in part by the Department of Energy through DE-FOA-0002563 and by AFOSR award FA9550-22-1-0098.
The work of YN was also supported in part by MEXT KAKENHI grant number JP20H05850, JP20H05860.

\appendix

\section{Avoided crossings}
\label{app:avoided}

Here we consider the avoided crossing
\begin{equation}
    G_{\Delta\,\Delta_w\,\text{nonpert}}(T_{1},T_{2})\supset   \begin{tikzpicture}[baseline={([yshift=-.5ex]current bounding box.center)}, scale=0.7]
        \pgftext{\includegraphics[scale=1]{fire20.pdf}} at (0,0);
        \draw (-2,0.7) node {$\color{red}\mathcal{O}_\Delta$};
        \draw (2,-0.7) node {$\color{red}\mathcal{O}_\Delta$};
        \draw (2.6,0.6) node {$\b_1$};
        \draw (-2.6,-0.6) node {$\b_3$};
        \draw (0,1.7) node {$\b_2$};
        \draw (0,-1.7) node {$\b_4$};
        \draw (-2.6,-1.4) node {$\color{blue}\mathcal{O}_{\Delta_w}$};
        \draw (2.6,1.4) node {$\color{blue}\mathcal{O}_{\Delta_w}$};
    \end{tikzpicture}\,\,.
\end{equation}
The exact amplitude is
\begin{equation}
  \frac{e^{-2\S}}{Z(\beta)}\int\! \d E\, e^{-\beta E}\,\rho(E)\,\int_{-\infty}^{+\infty}\!\d\ell\, \psi_{E}(\ell)\psi_{E}(\ell)\,e^{-\Delta \ell}\int_{-\infty}^{+\infty}\!\d\ell_w\, \psi_{E}(\ell_w)\psi_{E}(\ell_w)\,e^{-\Delta_w \ell_w}\,.
\end{equation}
At fixed energy, this simply factorizes
\begin{equation}
  e^{-2\S}\int_{-\infty}^{+\infty}\!\d\ell\, \psi_{E}(\ell)\psi_{E}(\ell)\,e^{-\Delta \ell}\int_{-\infty}^{+\infty}\!\d\ell_w\, \psi_{E}(\ell_w)\psi_{E}(\ell_w)\,e^{-\Delta_w \ell_w}\,.
\end{equation}
Each of these integrals computes the expectation value of a two-point function in the TFD at $t=0$, up to a factor of $1/\rho_0(E)$. This is clear from~\eqref{eq:2point-JT}. Alternatively, one can just do the $\ell$ and $\ell_w$ integrals, which results in gamma functions~\cite{Blommaert:2018oro}. Then, using Stirling's approximation for the gamma functions, one indeed recovers
\begin{equation}
  G_{\Delta\,\Delta_w\,\text{avoided}}(T_{1},T_{2})= \frac{1}{T_H^2}\,e^{\Delta \log(E)}\,e^{\Delta_w \log(E)}\,,\quad T_H=2\pi \rho(E)\,.
\end{equation}

Stripping off the normalization factor $e^{\Delta_w \log(E)}$ results in
\begin{equation}
  \mathcal{A}_\text{avoided}(T_1,T_2,\ell)=\frac{1}{T_H^2}\,\delta(\ell+\log(E))\,,
\label{eq:aa}
\end{equation}
therefore giving the contribution
\begin{equation}
  P_\text{safe}(T_1,T_2)\supset P_\text{avoided}(T_1,T_2) = \frac{1}{T_H^2}\,.
\label{eq:aa2}
\end{equation}
This is negligible in the tau-scaling limit. One should compare \eqref{eq:aa} with our main contribution to the amplitude due to wormholes that we studied in the main text~\eqref{eq:F-Fourier}. That amplitude too is suppressed by $1/T_H^2$. However, it has support for large ranges of $T_\text{eff}$ of order $T_H$, and the integral over such a large range results in a leading order correction to the safe/dangerous probabilities, unlike~\eqref{eq:aa} which has delta support on $T_\text{eff}=0$ resulting in a subleading contribution~\eqref{eq:aa2}.

\bibliographystyle{ourbst}
\bibliography{Refs}

\end{document}